\documentclass{aa}  

\usepackage{graphicx}
\usepackage[urlcolor=blue, colorlinks=true, citecolor=blue, linkcolor=blue]{hyperref}
\usepackage{txfonts}
\usepackage{placeins}
\usepackage{csquotes}
\usepackage[utf8]{inputenc}  
\usepackage[T1]{fontenc}    
\begin{document}

   \title{Automated quasar continuum estimation using neural networks}

   \subtitle{A comparative study of deep-learning architectures}

   \author{Francesco~Pistis\inst{\ref{unimib}, \ref{ncbj}, \ref{inaf_bologna}}
            \and Michele~Fumagalli\inst{\ref{unimib}, \ref{inaf_trieste}}
            \and Matteo~Fossati\inst{\ref{unimib}, \ref{inaf_brera}}
            \and Trystyn~Berg\inst{\ref{unimib}, \ref{nrc}, \ref{camosun}}
            \and Elena~S.~Mangola\inst{\ref{unimib}, \ref{marseille}}
          \and Rajeshwari~Dutta\inst{\ref{pune}}
          \and Margherita~Grespan\inst{\ref{ncbj}}
          \and Angela~Iovino\inst{\ref{inaf_brera}}
          \and Katarzyna~Ma\l{}ek\inst{\ref{ncbj}}
          \and Sean~Morrison\inst{\ref{illinois}}
          \and David~N.~A.~Murphy\inst{\ref{cambridge}, \ref{santiago}}
          \and William~J.~Pearson\inst{\ref{ncbj}}
          \and Ignasi~P\'{e}rez-R\'{a}fols\inst{\ref{pol_barc}}
          \and Matthew~M.~Pieri\inst{\ref{marseille}}
          \and Agnieszka~Pollo\inst{\ref{ncbj}, \ref{ju}}
          \and Daniela~Vergani\inst{\ref{inaf_bologna}}
          }

   \institute{Dipartimento di Fisica \enquote*{G. Occhialini}, Universit\`{a} degli Studi di Milano-Bicocca, Piazza della Scienza 3, I-20126 Milano, Italy\\
              \email{francesco.pistis@unimib.it}\label{unimib}
                \and 
                National Centre for Nuclear Research, ul. Pasteura 7, 02-093 Warsaw, Poland\label{ncbj}
                \and
                INAF - Osservatorio di Astrofisica e Scienza dello Spazio di Bologna, Via Piero Gobetti 93/3, I-40129 Bologna, Italy\label{inaf_bologna}
                \and
                INAF - Osservatorio Astronomico di Trieste, via G.B. Tiepolo 11, I-34143 Trieste, Italy\label{inaf_trieste}
                \and
                INAF - Osservatorio Astronomico di Brera, Via Brera 28, 20122 Milano, via E. Bianchi 46, 23807 Merate, Italy\label{inaf_brera}
                \and
                NRC Herzberg Astronomy and Astrophysics Research Centre, 5071 West Saanich Road, Victoria, B.C., Canada, V9E 2E7\label{nrc}
                \and
                Department of Physics and Astronomy Camosun College, 3100 Foul Bay Rd, Victoria, B.C., Canada, V8P 5J2\label{camosun}
              \and
              Aix Marseille Univ. CNRS, CNES, LAM, Marseille, France\label{marseille}
              \and
              IUCAA, Postbag 4, Ganeshkind, Pune, 411007, India\label{pune}
              \and
              Department of Astronomy, University of Illinois at Urbana-Champaign, Urbana, IL 61801, USA\label{illinois}
              \and
              Institute of Astronomy, University of Cambridge, Madingley Road, Cambridge CB3 0HA, UK\label{cambridge}
              \and
              Instituto de Astrofisica, Facultad de Fisica, Pontificia Universidad Catolica de Chile, Santiago, Chile\label{santiago}
              \and
              Departament de F\'{i}sica, EEBE, Universitat Polit\'{e}cnica de Catalunya, c/Eduard Maristany 10, 08930 Barcelona, Spain\label{pol_barc}
              \and
              Astronomical Observatory of the Jagiellonian University, Orla 171, 30-001 Cracow, Poland\label{ju}
              }

   \date{Received ; accepted }

 
  \abstract
   {Ongoing and upcoming large spectroscopic surveys
   are drastically increasing the number of observed quasar spectra, requiring the development of fast and accurate automated methods to estimate spectral continua.}
   {This study evaluates the performance of three neural networks (NN) --- an autoencoder, a convolutional NN (CNN), and a U-Net --- in predicting quasar continua within the rest-frame wavelength range of $1020~\text{\AA}$ to $2000~\text{\AA}$. The ability to generalize and predict galaxy continua within the range of $3500~\text{\AA}$ to $5500~\text{\AA}$ is also tested.}
   {The performance of these architectures 
   is evaluated using the absolute fractional flux error (AFFE) on a library of mock quasar spectra for the WEAVE survey, and on real data from the Early Data Release observations of the Dark Energy Spectroscopic Instrument (DESI) and the VIMOS Public Extragalactic Redshift Survey (VIPERS).
   }
   {The autoencoder outperforms the U-Net, achieving a median AFFE of 0.009 for quasars. The best model also effectively recovers the Ly$\alpha$ optical depth evolution in DESI quasar spectra.
   With minimal optimization, the same architectures can be generalized to the galaxy case, with the autoencoder reaching a median AFFE of
   0.014 and reproducing the D4000n break in DESI and VIPERS galaxies. 
   }
   {}


   \maketitle
%

\section{Introduction}\label{sec:intro}

Quasars are important probes of the distribution of baryonic matter, which is imprinted on the spectra in the form of absorption lines.
For example, the distribution of neutral hydrogen in the intergalactic medium (IGM) is observed as a dense forest of absorbers at wavelengths lower than $1\,216\,\text{\AA}$ in the quasar rest-frame, the so-called Lyman-$\alpha$ forest \citep[Ly$\alpha$,][]{lynds1971abs, bahcall1971abs, mcdonald2006lyalpha, shull2012baryon}.
Hydrodynamical simulations contributed to the interpretation of the Ly$\alpha$ forest, showing that hydrogen follows the underlying dark matter distribution in the network of filaments, voids, and clusters \citep{cen1994lyalpha, hernquist1996lyalpha, croft1998lyalpha}.
At the same time, at wavelengths $\lambda >1216$~\AA~in the quasar rest frame, spectroscopy offers a detailed view of the composition and kinematics of the diffuse, metal-enriched gas \citep[through the measurements of absorption lines such as \ion{C}{iv}, \ion{Mg}{II}, and \ion{Si}{iv};][]{dutta2020magg, galbiati2023magg} in between and near galaxies, within the circumgalactic medium (CGM).
High-quality estimates for the quasar continua are required to measure the optical depth of the transitions producing the absorption lines and hence extract physical information from these datasets.

Likewise, galaxy spectra provide critical information about how the gas in the CGM is eventually converted and processed within the interstellar medium (ISM) and stars. These transformations are imprinted in relations such as the mass-metallicity relation \citep{erb2010mzr, wuyts2012mzr, zahid2014mzr, bian2017mzr} or the fundamental metallicity relation \citep{mannucci2010fundamental, cresci2019fundamental, curti2020mass, pistis2022bias, pistis2024cross}.
To have a reliable measurement of the absorption and emission features (e.g., the metallicity derived as the relative abundance of oxygen to hydrogen), it is, once again, essential to accurately determine the continuum of the galaxy spectra. Moreover, the continua of galaxies contain a plethora of information about the stellar population \citep{maraston2005stellarpop, maraston2011stellar, vazdekis2010miles}.

The increasing amount of data since the advent of big spectroscopic surveys such as the Sloan Digital Sky Survey \citep[SDSS,][]{abazajian2009sdss}, the VIMOS Public Extragalactic Redshift Survey \citep[VIPERS,][]{scodeggio2018vimos}, the Dark Energy Spectroscopic Instrument \citep{desi2022overview}, and the ESA Euclid mission \citep{laureijs2011euclid}, or the upcoming WHT Enhanced Area Velocity Explorer \citep[WEAVE,][]{dalton2012weave, dalton2014weave, dalton2016weave, pieri2016weaveqso, jin2024weave} and the 4-meter Multi-Object Spectroscopic Telescope \citep[4MOST,][]{dejong20194most}, make the estimate of continuum spectra an untractable problem with traditional continuum fitting techniques based on, e.g., minimization of the $\chi^2$ or via fitting of stellar population synthesis (SPS) models.
Thus, it is necessary to resort to fully automatic algorithms suitable for accurate and precise measurements in large catalogs using a limited amount of computational time and resources.

In this context, machine learning (ML) has emerged as a powerful tool to approach the challenges of continuum fitting.
Various ML algorithms have been developed with different applications to astrophysical spectra including: classification of the sources \citep{folkes1996class, geach2012class, pat2022class, wang2023class, abraham2024class, wu2024class};  redshift measurements \citep{machado2013redshift, giri2020redshift};  analysis of lines both in emission \citep{begue2024blazar} and absorption \citep{jalan2024abs}, or stellar population \citep{liewcain2021stellar, murata2022spectralfitting, wang2024stellar}.
ML algorithms can automatically learn complex patterns from large amounts of data, making them ideal for processing the extensive data sets generated by modern spectroscopic surveys. 
For example, neural networks (NNs) and other ML techniques can be trained to predict the quasar continuum by learning from a substantial number of spectra. 
This approach can accommodate the broad emission lines and the intricate shape of the quasar continuum on the red and blue sides of the Ly$\alpha$ emission \citep{paris2011pca, greig2017lyalpha}, as well as the complexity of galaxy spectra \citep{portillo2020dimred, teimoorinia2023galdiv, melchior2023spender, bohm2023anomalydet, liang2023outdet, liang2023aegal}.

Considering quasars, which is the main focus of this study, early explorations of this approach include the prediction of the true continuum using only the red part of the spectra (redward of the Ly$\alpha$ line) using principal component analysis \citep[PCA,][]{suzuki2005qso, paris2011pca, lee2012pca, davies2018pca}. However, additional constraints on the Ly$\alpha$ mean flux \citep{lee2012pca} are often required to improve the predictions in the forest region.
ML models offer a more flexible and potentially more accurate alternative by directly learning the relationship between different spectral regions and the underlying continuum, reducing the need for such corrections.
Recently, efforts have moved toward a deep-learning approach.
For example, a feedforward NN (such as an autoencoder) shows good results predicting the quasar continuum in the Ly$\alpha$ forest region using as input only information on the red side of the spectra \citep{liu2021quasar}.
Another approach is to use a convolutional NN (CNN). This kind of architecture shows, on average, lower errors than the feedforward NN \citep{turner2024lycan}.
A similar direction has been taken for studying galaxy spectra. 
Various approaches have been applied for estimating physical properties via supervised algorithms from spectro-photometric properties \citep[for example,][]{angthopo2024weaveml}, directly applied to the spectrum to predict physical quantities of emission lines \citep[for example,][]{ucci2017game, ucci2018game, ucci2019game} or morphological classification \citep[for example,][]{vavilova2021morphml}.

In this work, we perform a comparative analysis of various architectures applied to the problem of estimating the quasar continuum. Specifically, we test an autoencoder \citep[of the type used by][]{liu2021quasar}, a CNN \citep[of the type used by][]{turner2024lycan}, and a U-Net \citep{ronneberger2015unet}.
These NNs are trained on WEAVE mock quasar spectra and tested for generalization on the early data release (EDR) of DESI.
Starting from models designed for the analysis of quasars, we further test the ability of these algorithms to generalize to different continuum shapes with minimal optimizations. This is achieved by studying the performance of the autoencoder, the CNN, and the U-Net architecture on galaxy spectra from the VIMOS Public Extragalactic Redshift Survey (VIPERS) survey.

The paper is organized as follows.
In Sect.~\ref{sec:data} we describe the primary samples used in this work, in Sect.~\ref{sec:ml} we outline the ML approach and data preprocessing, while in Sect.~\ref{sec:application}, we detail the application of the ML algorithm to quasars and its generalization test to different spectral shapes such as galaxies.
In Sect.~\ref{sec:phys_test} we describe the applicability of these algorithms to different unseen datasets, using DESI data.
In Sect.~\ref{sec:eff} we describe our considerations on computational efficiency.
In Sect.~\ref{sec:gal} we test our NNs with minimum tuning using galaxy spectra.
Finally, we summarize our results in Sect.~\ref{sec:sum}.

\section{Data samples}\label{sec:data}

Here, we describe the three main samples used in this study. The quasar samples are based on mock observations of the WEAVE survey (see Sect.~\ref{subsec:weave}) and real observations from the DESI EDR survey (see Sect.~\ref{subsec:desi}).
The galaxy sample is based on real observations of the VIPERS survey (see Sect.~\ref{subsec:vipers}) and DESI EDR survey (see Sect.~\ref{subsec:desi}).

\begin{figure*}
    \centering
    \resizebox{\hsize}{!}{\includegraphics{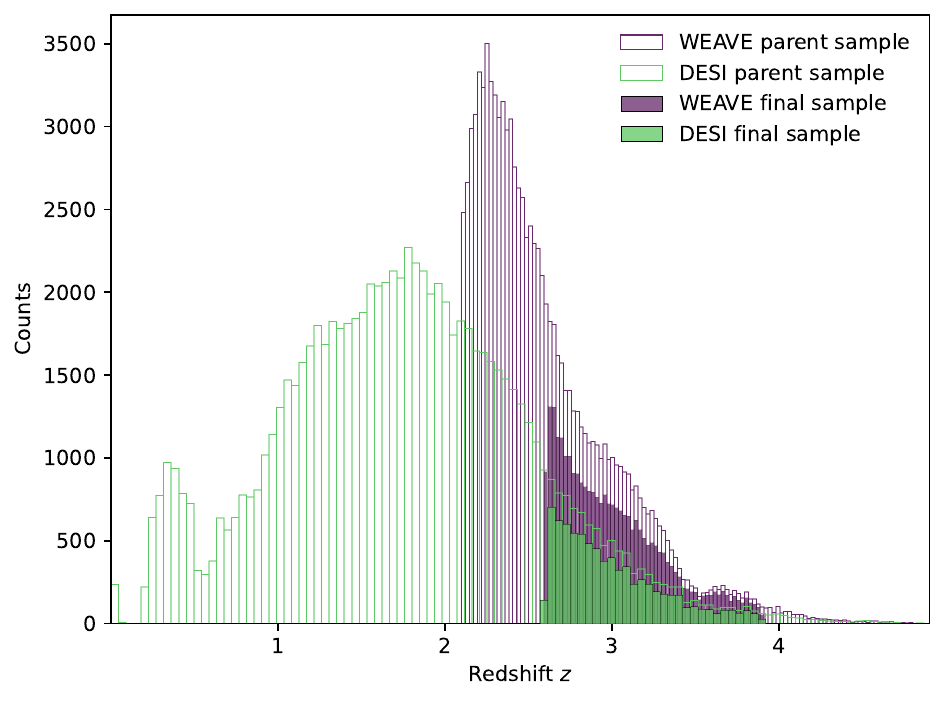}\includegraphics{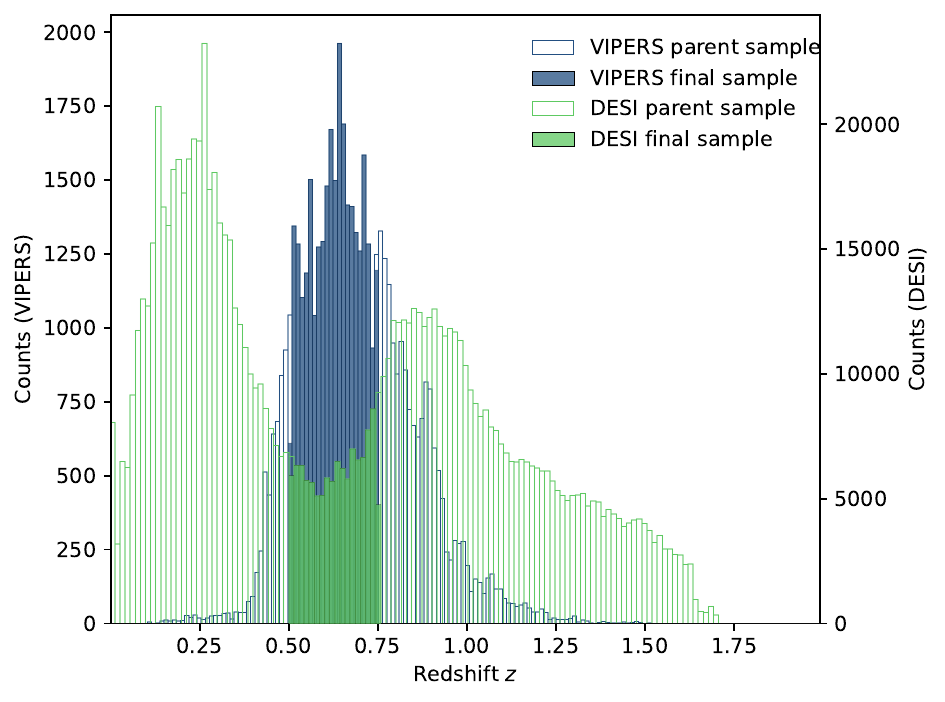}}
    \resizebox{\hsize}{!}{\includegraphics{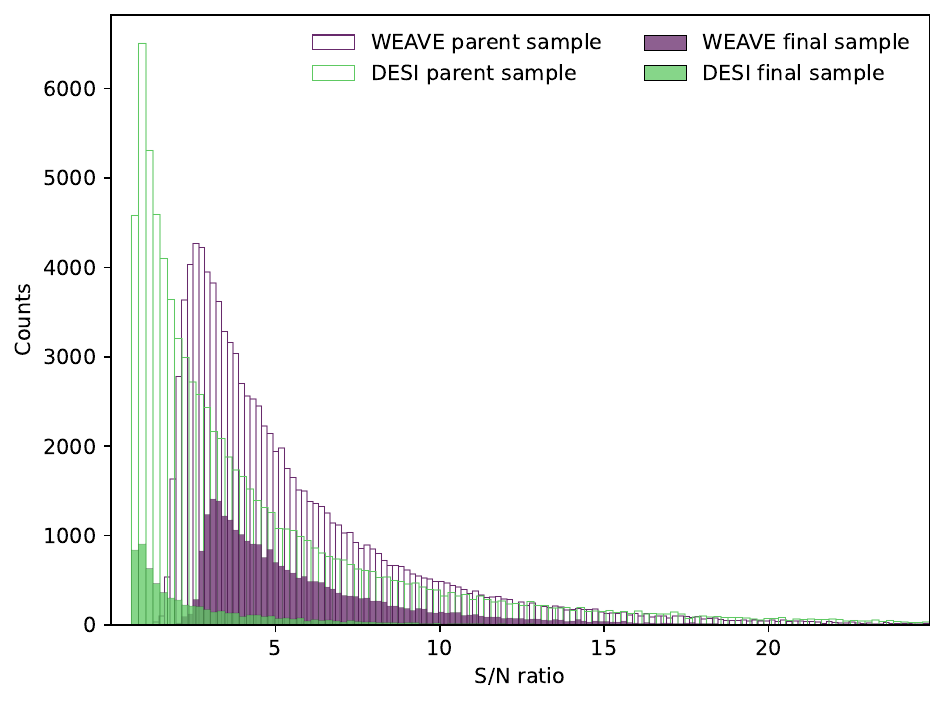}\includegraphics{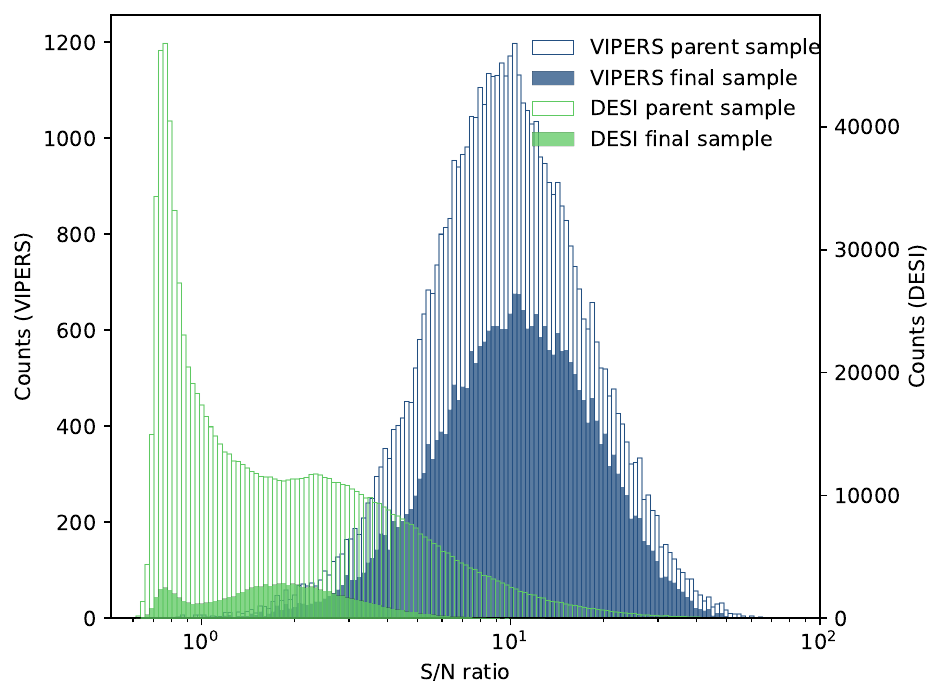}}
    \caption{Histograms of the redshift (top panel) and $\mathrm{S/N}$ (bottom panel) for quasars (left column in purple for the WEAVE mock catalog and in green for DESI EDR) and galaxies (right panel in blue for the VIPERS and in green for DESI EDR). The unfilled histograms show the distributions of the parent samples, while the filled histograms show the distributions of the final samples after all the data selections.}
    \label{fig:dist_zspec}
\end{figure*}

\subsection{WEAVE quasar mock catalog}\label{subsec:weave}

WEAVE is a next-generation wide-field spectroscopic survey at the $4.2$-m WHT at the
Observatorio del Roque de los Muchachos on La Palma, Spain \citep{dalton2012weave, dalton2016weave}.
The facility includes a $2$-deg field-of-view focus corrector system with a $1\,000$-multiplex fiber positioner.
The science focus of the main surveys to be performed using WEAVE includes the study of stellar and galaxy evolution in different environments over the last $5$--$8$ Gyr and the changing scale of the Universe \citep{jin2024weave}.
Among the science cases of WEAVE, WEAVE-QSO \citep[WQ,][]{pieri2016weaveqso}, WEAVE Galaxy Clusters Survey, and WEAVE Stellar Populations at intermediate redshifts Survey \citep[StePS,][]{iovino2023weavesteps} are of interest for this work.

WQ will observe around $\sim 450\,000$ quasars at high redshift over an area of $\sim 10\,000~\mathrm{deg}^2$  \citep[termed WQ-Wide;][]{jin2024weave}.
All WQ targets are chosen to obtain quasar spectra with $z_\mathrm{q} > 2.2$.
This limit allows the coverage of the Ly$\alpha$ forest at $z>2$.
WEAVE GalaXy Clusters Survey will observe around $\sim 200\,000$ galaxies at low redshift ($z < 0.5$) over an area of $\sim 1\,350~\mathrm{deg}^2$ \citep{jin2024weave}. The WEAVE-StePS, instead, will observe around $25\,000$ galaxies at intermediate redshift ($0.3 < z < 0.7$) over an area of $\sim 25~\mathrm{deg}^2$ \citep{jin2024weave}.

Because data are not yet available, we used a library of $100\,000$ simulated quasar spectra compiled by drawing random continuum shapes from a set of principal components derived from observed high-quality BOSS spectra \citep{paris2011pca, paris2012dr9, paris2014dr10, paris2017dr12, paris2018dr14}. These mock spectra also contain a model for the Ly$\alpha$ forest based on skewers extracted from hydrodynamic simulations \citep{bolton2017sherwood}. Intervening metal absorption line systems associated with quasars are also included to increase the realism of these mocks, with a frequency and a redshift distribution tuned to reproduce the observed statistics \citep{paris2018dr14, shu2019agngaia, lyke2020dr16sdss}. These idealized spectra are subsequently resampled and convolved with a line-spread function model to match the resolution $R$ of the WEAVE survey, $R = 5\,000$ and $R = 20\,000$ for low- and high-resolution modes, respectively, for wavelength coverage over the range $366$--$959$ nm \citep{jin2023weave}. Flux-dependent noise is also included to simulate WEAVE-like observations (at the observational condition of air mass $a=1.107$ and apparent magnitude of the sky $b_\mathrm{sky}=20.92\ \text{mag}/\text{arcsec}^2$), although the presence of skylines is currently not included. 

For this catalog, we apply two main selections.
The first is the selection of the magnitude, keeping all sources with $r$-band magnitude lower than 21 (see Appendix~\ref{app:ffe_bias} for analysis of the bias introduced by different cuts).
The second selection is in the redshift range, limited to $2.6 \leq z \leq 3.9$ to have the entire rest-frame wavelength range of interest $1\,020~\text{\AA}$--$2\,000~\text{\AA}$ covered in the data.
We also checked the effects of a selection in S/N of the spectra on the quality of the continuum fit. We found minor to no effects of the selection on S/N (see Appendix~\ref{app:snr} for details).
These selections reduce the sample size to $\sim 30\,000$ mock spectra.
Fig.~\ref{fig:dist_zspec} shows the redshift and $\mathrm{S/N}$ distributions of the WEAVE quasar mock catalog and, for comparison, the DESI/EDR catalog (Sect.~\ref{subsec:desi}).
The WEAVE mock catalog reproduces well the redshift distribution of DESI quasars in the range considered.

\subsection{VIPERS galaxy catalog}\label{subsec:vipers}

The main goal of this analysis is to compare the performances of various ML techniques in predicting the continuum of quasars. As part of this study, we test the ability of the selected algorithms to predict different spectral shapes, including those of galaxies. For this, we apply the same methodology used on quasars also on galaxy spectra.
To compile a galaxy sample, we use the catalog from the VIMOS Public Extragalactic Redshift Survey \citep[VIPERS,][]{guzzo2013vipers, garilli2014vimos, scodeggio2018vimos} that is a spectroscopic survey completed with the VIMOS spectrograph \citep{le2003proc}. 
Its primary purpose was to measure the redshift of almost $10^5$ galaxies in the $0.5 < z <1.2$ range. The area covered by VIPERS is about $25.5~\text{deg}^2$ on the sky. Only galaxies brighter than $i_{AB} = 22.5$ were observed relying on a preselection in the $(u-g)$ and $(r-i)$ color-color plane to remove galaxies at lower redshifts (see \citealt{guzzo2014vimos} for a more detailed description). 
The Canada-France-Hawaii Telescope Legacy Survey Wide \citep[CFHTLS-Wide,][]{mellier2008cfhtls} W1 and W4 equatorial fields compose the galaxy sample, at $\text{R.A} \simeq 2$ and $\simeq 22~\text{hours}$, respectively.
The VIPERS spectral resolution ($R \sim 250$) allows us to study individual spectroscopic properties of galaxies with an observed wavelength coverage of $5\,500$--$9\,500~\text{\AA}$.
The data reduction pipeline and the redshift quality system are described in \cite{garilli2014vimos}.
The final data release provides reliable spectroscopic measurements and photometric properties for $86\,775$ galaxies \citep{scodeggio2018vimos}.

Starting from the VIPERS Public Data Release-2 (PDR2) spectroscopic catalog,
we limit our sample to galaxies with a redshift confidence level of $99$~per cent ($3.0 \leq z_\text{flag} \leq 4.5$), reducing the redshift interval of the catalog within the range $0.4 < z < 1.2$ (see Fig.~\ref{fig:dist_zspec} for the redshift and $\mathrm{S/N}$ distributions).
This selection reduces the sample to $\sim 50\,000$ sources, including narrow-line AGNs, star-forming galaxies, and passive galaxies.
The selection on the redshift confidence level interval has the side effect of shifting the distribution in $\mathrm{S/N}$ toward higher values. Galaxies with the chosen interval of $z_\text{flag}$ show apparent spectral features in the spectrum used to measure the redshift.
The second selection is in the redshift range, limited to $0.5 \leq z \leq 0.75$ to have the entire rest-frame wavelength range of interest $3\,500~\text{\AA}$--$5\,500~\text{\AA}$ covered in the data.
These selections reduce the sample size to the $\sim 30\,000$ observed galaxy spectra.

Having a good measurement of the true continuum as a label during the training of the NN model is essential.
We estimate the continuum of the VIPERS galaxy spectra by modeling them \citep[as done in][]{pistis2024cross} using the penalized pixel fitting code \citep[pPXF,][]{cappellari2004fit, cappellari2017improving, cappellari2022full}, which makes it possible to fit both the stellar and gas components via full-spectrum fitting.
After shifting the observed spectra to the rest frame and masking out the emission lines, the stellar component of the spectra is fitted with a linear combination of stellar templates from the MILES library \citep{vazdekis2010miles} convolved to the same spectral resolution as the observations.
The gas component is fitted with a single Gaussian for each emission line.

\begin{figure}
    \centering
    \resizebox{\hsize}{!}{\includegraphics{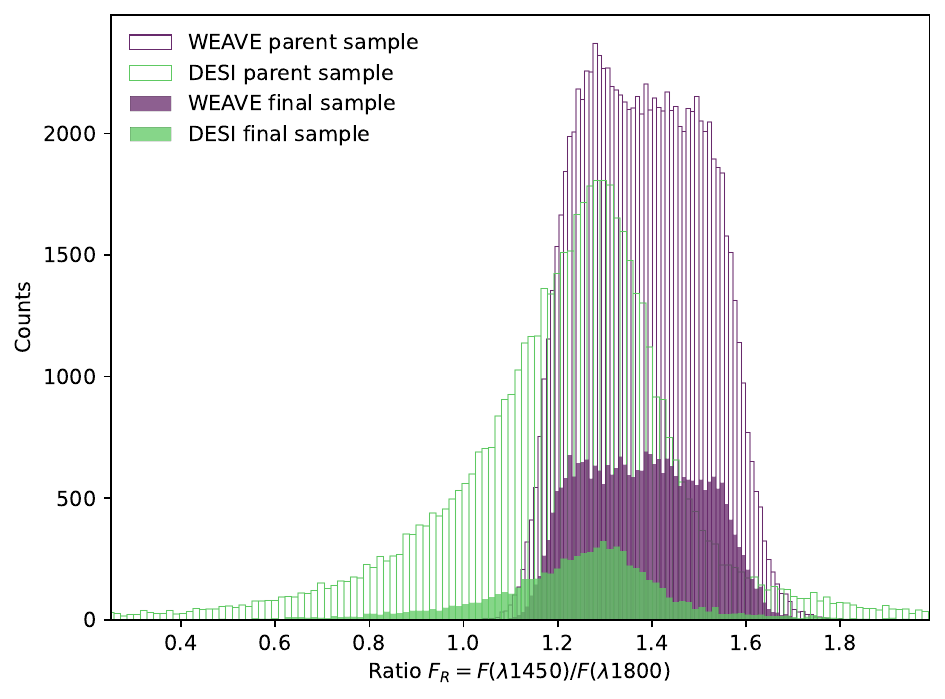}}
    \resizebox{\hsize}{!}{\includegraphics{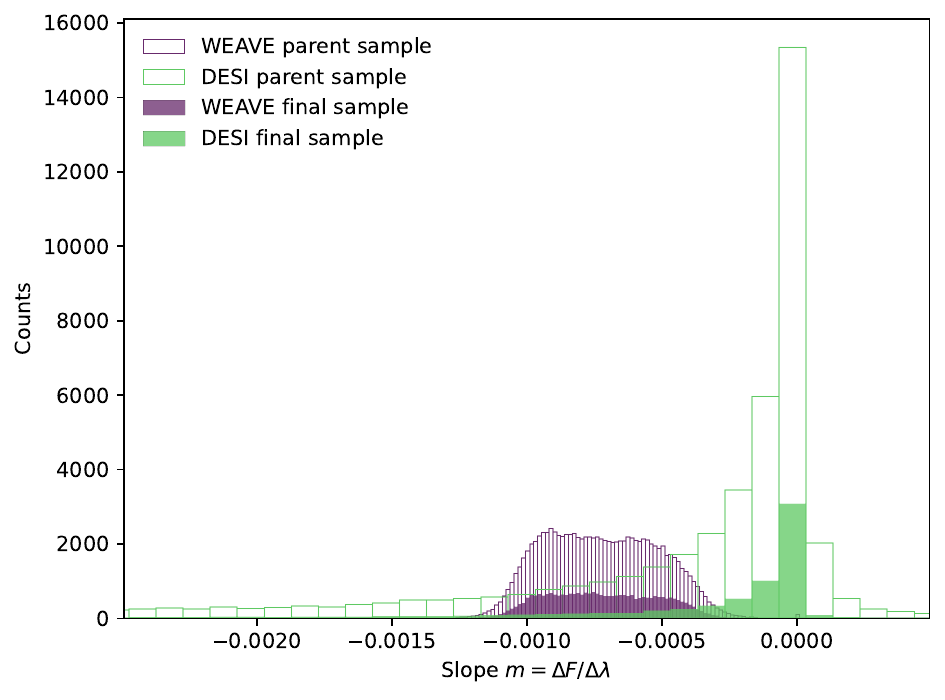}}
    \caption{Histograms of the ratio (top) and slope (bottom) of the quasar spectra between the rest-frame wavelengths $\lambda = 1800$~\AA~and $\lambda = 1450$~\AA~for the WEAVE mock catalog (purple) and the DESI EDR (green). The unfilled histograms show the distributions of the parent samples, while the filled histograms show the distributions of the final samples after all the data selections.}
    \label{fig:dist_ratio_slope}
\end{figure}

\subsection{DESI quasar and galaxy catalogs}\label{subsec:desi}

While the WEAVE mock and VIPERS catalogs are used to train and test the NN models, we also studied how these models can be deployed on unseen datasets with similar properties.
For this step, we chose the Early Data Release\footnote{\url{https://data.desi.lbl.gov/doc/}} (EDR) of the Dark Energy Spectroscopic Instrument \citep[DESI,][]{desi2022overview, desi2023edr} which is the largest spectroscopic survey to date with characteristics similar to those expected for WEAVE, such as the spectroscopic resolution (between $2\,000$ and $5\,000$ at different wavelength intervals) and the redshift range of the observed targets.
Galaxies in DESI/EDR also share redshifts similar to those of the VIPERS sample.
For quasars, we removed the broad absorption line (BAL) quasars.
The BALs are removed because the mock catalog at present does not include these sources and their presence would lead to a more complex assessment of the generalization performance without having a close look at these specific cases.
For this task, we use the public value added catalog\footnote{\url{https://data.desi.lbl.gov/doc/releases/edr/vac/balqso/}} (VAC) by \cite{filbert2023baldesi}.
The VAC catalog contains the absorption indices \citep[AI, defined by][]{hall2002abs} for the metal lines \ion{C}{IV} and \ion{Si}{IV}.
We considered a BAL any quasar with an AI value higher than 0 for any metal line.
For both quasars and galaxies, we selected sources with redshift ranges that cover the same interval as the WEAVE mock and VIPERS catalogs.
The DESI catalogs are reduced to $9\,000$ quasars and $100\,000$ galaxies (see Fig.~\ref{fig:dist_zspec} for the redshift and $\mathrm{S/N}$ distributions).

The public DESI EDR catalog contains the continua of both quasars and galaxies computed with Redrock, which is intended as a redshift fitter and classifier. Redrock fits PCA templates of three broad, independent classes of objects: stars, galaxies, and quasars.
The spectra is classified according to its minimum $\chi^2$ of the fit.
In the following, we will compare our results against the available Redrock continuum, which is independent of what is obtained through NNs.
However, the DESI pipeline has been expanded, for example, with QuasarNET \citep{busca2021quasarnet} for quasars and SPENDER \citep{melchior2023spender, liang2023outdet, liang2023aegal} for galaxies, which provide NN-based high-quality continua for DESI sources.

Before proceeding with our work, we verify that DESI and the WEAVE mock quasars share broadly similar continuum shapes. For this, we compared the distribution of the flux ratio $F_R$, defined as $F_R=F(\lambda = 1450 ~\text{\AA} ) / F(\lambda = 1800 ~\text{\AA})$, and the slope $m$, defined as $\Delta F / \Delta \lambda$ with $\Delta \lambda = 1800-1450~\text{\AA}$. This comparison is shown in Fig.~\ref{fig:dist_ratio_slope}. 
WEAVE mocks generally overlap with the DESI data for the $R$ parameter, while displaying a slope distribution shifted toward lower values. As we show in the following sections, the best models trained on WEAVE data generalize sufficiently well and can be applied to DESI data with satisfactory success.

Finally, we employ the DESI sample to test our NN's performances in recovering physical quantities. For the quasar case, we estimate the evolution of the mean optical depth (see Sect.~\ref{sec:phys_test} for details) and compare with previous results \citep{becker2013tau, turner2024lycan}.
For the galaxy case, we compute the D4000n break and compare the results between the NNs and Redrock\footnote{\url{https://github.com/desihub/redrock}}, released with the DESI EDR. We did not compute the underlying continua for the DESI spectra with other traditional methods, such as pPXF used for VIPERS. 

\section{Machine learning approach and architectures}\label{sec:ml}

\begin{figure}
    \centering
    \resizebox{\hsize}{!}{\includegraphics{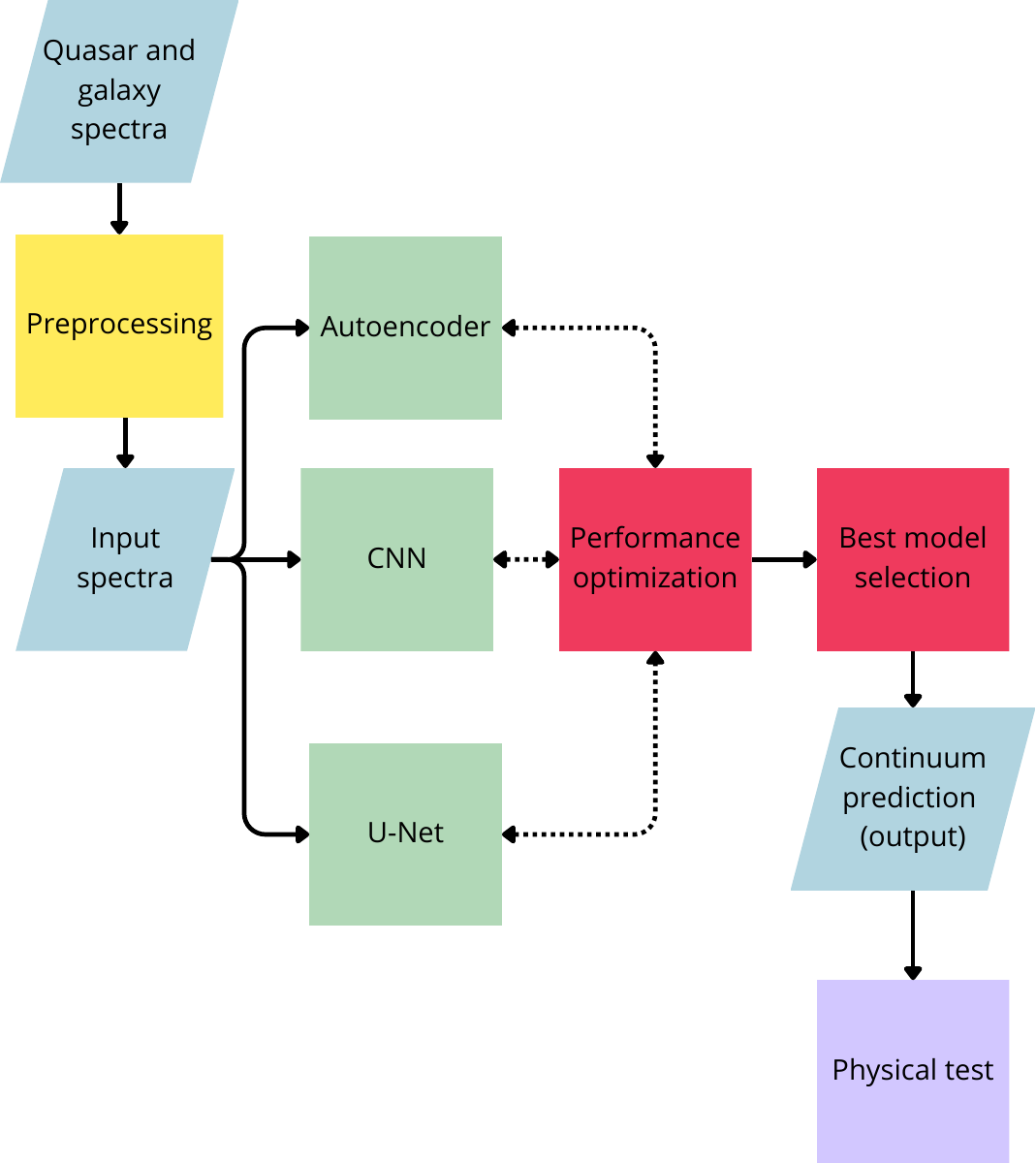}}
    \caption{Graphical representation of the workflow. The dotted double arrows represent recursive steps.}
    \label{fig:flow}
\end{figure}

We focus first on the reconstruction of the continuum for quasar spectra (see Sect.~\ref{subsec:app_qso}), and then we test how well the same architecture performs in modeling the stellar component in galaxy spectra after a retraining and optimization step (see Sect.~\ref{sec:gal}).
For this reason, quasars and galaxies are treated separately and each NN is optimized and trained independently between quasars and galaxies.
In this work, we used three different classes of NNs: autoencoders, CNNs, and U-Net.
Figure~\ref{fig:flow} shows the flow chart of the methodology followed in this work.
The raw spectra are preprocessed (see Sect.~\ref{subsec:preproc} for details) to produce similar spectra (in wavelength and flux ranges) to pass as input to the NNs.
These architectures are then optimized by randomly searching the hyperparameters (for example, the number and size of layers or kernels).
The optimization is performed with a random search for autoencoders for 100 trials. The CNNs and U-Net are optimized with a Bayesian search for 50 trials.
After optimizing each architecture, we selected the best model to predict the continua of the spectra.
The performances of each NN are then compared (see Sect.~\ref{sec:application} for details) via the absolute fractional flux error (AFFE) metric and bias in the prediction when found. 
The best architecture is finally tested on the physical aspect (see Sect.~\ref{sec:phys_test} for details).

During the training phase, we adopted a general strategy shared among all NNs.
We randomly divided the samples into training ($50$ per cent of the total sample), validation ($25$ per cent of the total sample), and test ($25$ per cent of the total sample).
At each training cycle, the order of the training spectra is randomly shuffled.
We adopt an early stopping strategy \citep{prechelt2012earlystopping} if there are no updates on the loss for the validation sample during the training process for five consecutive epochs.
We chose a loss defined by the mean squared error between the true and predicted continua.
Because the analysis is performed in the source's rest frame, not all spectra cover the whole wavelength range.
A few pixels can be missing at the edges of the spectra.
For this reason, we assign the value 0 to the missing parts at the edges of the spectra,  adding a masking layer to the autoencoder to exclude this portion of the spectra.

\subsection{Performance metrics}\label{subsec:metric}

To measure the quality of the fit, we use the AFFE as a fit-goodness metric (see, e.g., \citealt{liu2021quasar, turner2024lycan}).
The AFFE is defined as:
\begin{equation}
    \mathrm{AFFE} = \left| \delta F \right| = \left. \int_{\lambda_1}^{\lambda_2} \left| \frac{F_\mathrm{pred}\left( \lambda \right) - F_\mathrm{true}\left( \lambda \right)}{F_\mathrm{true}\left( \lambda \right)} \right|\, d\lambda \middle/ \int_{\lambda_1}^{\lambda_2} d\lambda \right.,
\end{equation}
where $F_\mathrm{pred}$ is the predicted output and $F_\mathrm{true}$ is the true continuum of the simulated quasar spectra or the pPXF fit of the stellar component for the galaxy spectra.
To check the presence of bias as a function of the wavelength, we use the fractional flux error (FFE) defined as:
\begin{equation}
    \mathrm{FFE} =\delta F \left( \lambda \right)  =  \frac{F_\mathrm{pred}\left( \lambda \right) - F_\mathrm{true}\left( \lambda \right)}{F_\mathrm{true}\left( \lambda \right)}.
\end{equation}

\subsection{Data pre-processing}\label{subsec:preproc}

Data preprocessing is a critical step in many ML applications.
This study follows a similar preprocessing used in \cite{turner2024lycan}.
The first step in predicting the continuum is to sample the spectra on the same grid in the same rest-frame range.
We interpolated the simulated quasar spectra in the range $1020$--$2000~\text{\AA}$ within pixels of $0.2~\text{\AA}$ width, while we interpolated the observed galaxy spectra in the NUV-optical range $3500$--$5500~\text{\AA}$ within pixels of $1~\text{\AA}$ width.
The difference is due to the disparity in spectral resolution between the WEAVE and VIPERS surveys.

The second step is to scale all the data to the same range to avoid the specific features of some spectra dominating over others.
Each quasar spectrum, both in the WEAVE and the DESI samples, is normalized by the median flux at $\lambda = 1450~\text{\AA}$ within a window of $\Delta \lambda = 50~\text{\AA}$.
The galaxy spectra, instead, are normalized according to the median flux over the whole wavelength range, as was done in the analysis with pPXF.

\begin{figure*}
    \centering
    \resizebox{\hsize}{!}{\includegraphics{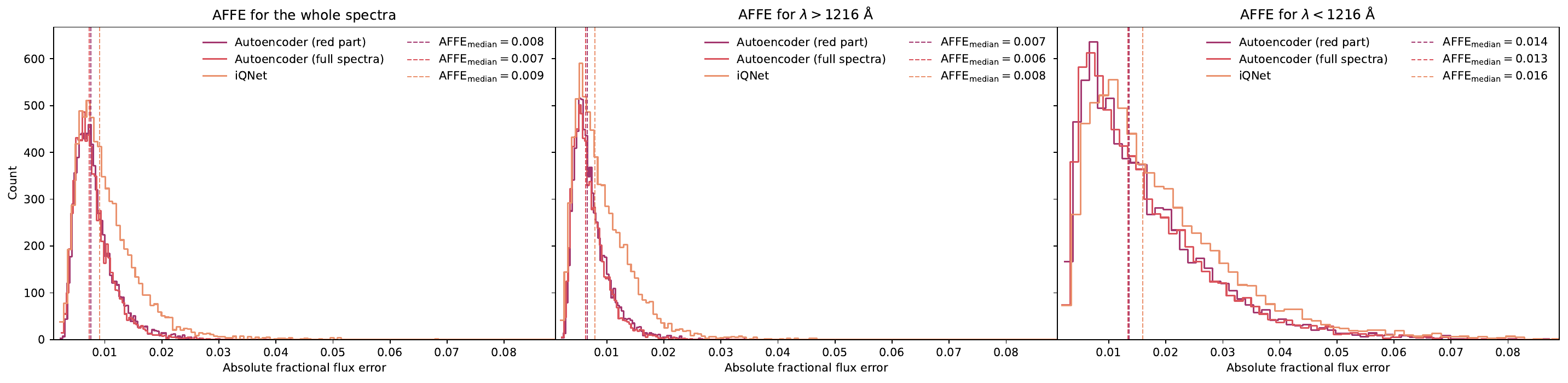}}
    \resizebox{\hsize}{!}{\includegraphics{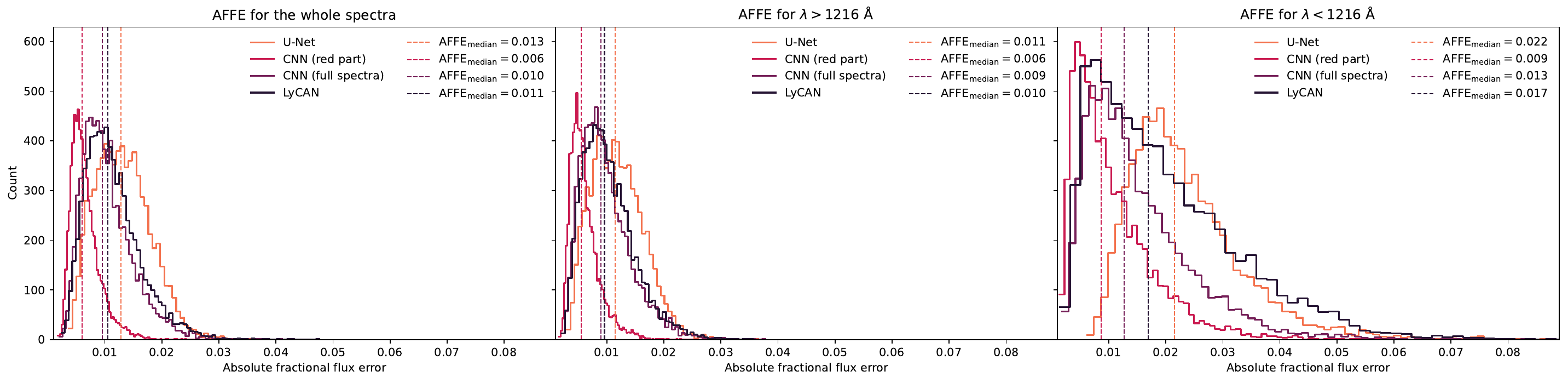}}
    \caption{Performance of NN predictions: histograms of the absolute fractional flux error (AFFE) for the autoencoders (top row) and CNN plus U-Net (bottom row) applied to the quasar datasets. The vertical lines show the median AFFE value for each histogram.}
    \label{fig:dist_affe_weave}
\end{figure*}

\begin{figure*}
    \centering
    \resizebox{\hsize}{!}{\includegraphics{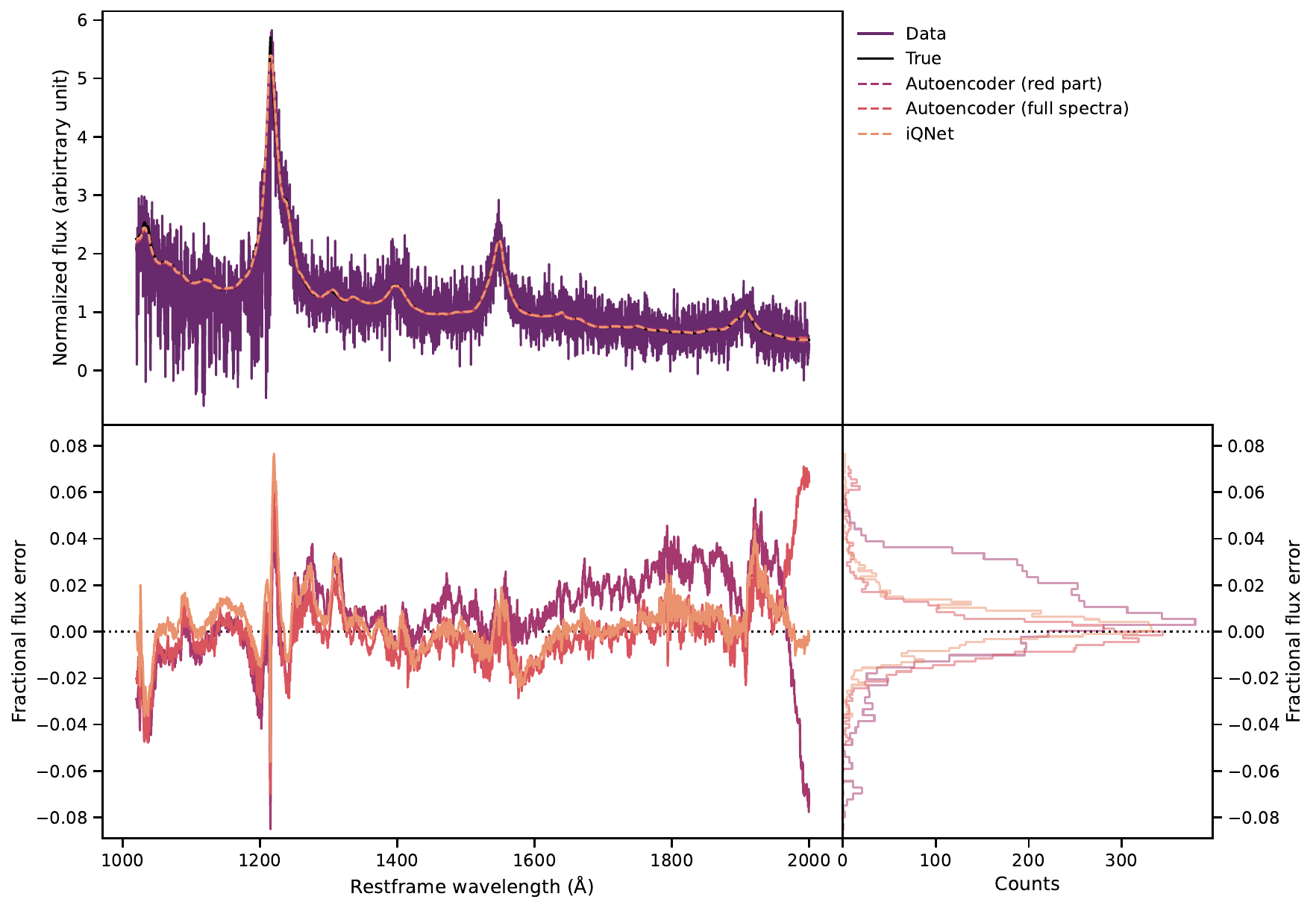}\includegraphics{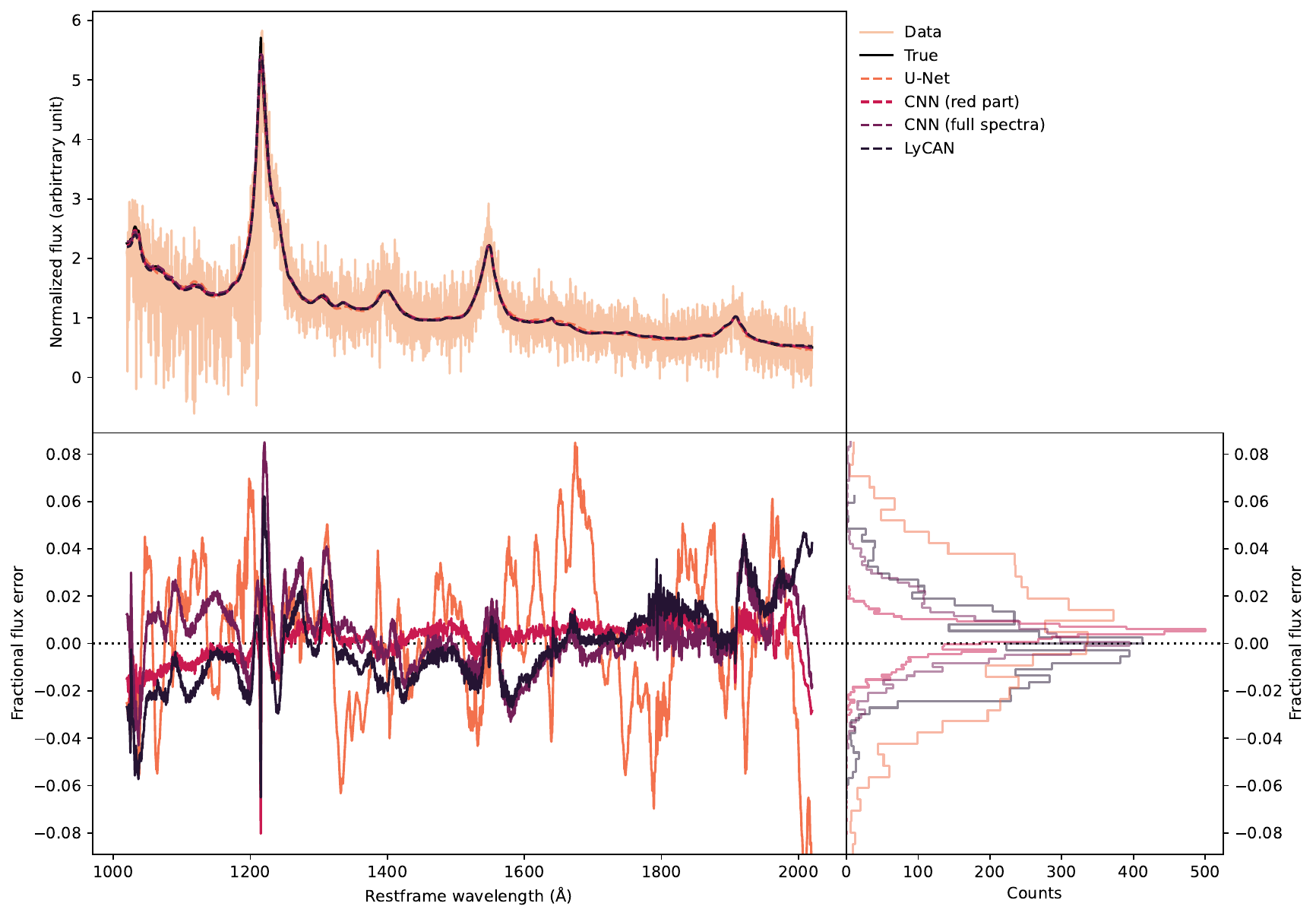}}
    \caption{Example quasar spectrum. The top panel of each plot shows the data, the true continuum, and the fits of different runs of the autoencoders (left) and CNNs (right). The bottom panel shows the FFE as a function of the rest-frame wavelength for different runs of the autoencoders and their distributions.}
    \label{fig:fit_weave}
\end{figure*}

\subsection{Autoencoders}\label{subsec:autoencoder}

The first architecture we consider in our study is the autoencoder.
They consist of an encoder and a decoder, where the encoder maps the input data into a lower-dimensional latent space, and the decoder attempts to reconstruct the original input from this compressed representation.
By minimizing the difference between the original data and its reconstruction, autoencoders learn to capture the most essential features of the input.
Our application relies on a general architecture where the encoder and decoder consist of three layers plus a layer with a bottleneck function. This is motivated by the intelligent quasar continuum neural network \citep[iQNet,][]{liu2021quasar}, which we take as a starting point. Further updates are performed to adapt the original IQNet for our specific problem, and the models we used for the different data are reported in Table~\ref{tab:iqnet} in the Appendix~\ref{app:arc}.

We also compare the modified architecture with the direct application of the original IQNet model, without any particular change except for the pre-processing of the data.
Specifically, normalizing the spectra by flux at $\lambda = 1450~\text{\AA}$ instead of the min-max scaling used by \cite{liu2021quasar} required us to change the loss function from binary cross-entropy \citep{liu2021quasar} to mean squared error.

\subsection{Convolutional neural networks}

We further explore the performance of a CNN starting from the Ly$\alpha$ Continuum Analysis Network \citep[LyCAN,][]{turner2024lycan}, which was designed to handle DESI spectra and which we use as a reference. 
A CNN is a regularized type of feedforward NN that learns features by itself via the application of a kernel or filter \citep{lecun2015deep}.
The kernel moves along the spectra with a given step, called a stride, performing an element-wise multiplication operation between the kernel and the input.
After each convolutional layer, a CNN has a pooling layer with the goal of reducing the size of the convolved feature.
The pooling layer can be of mainly two types: i) max-pooling where the max value within the kernel is taken; and ii) average-pooling where the average of all values within the kernel is taken.

We built our CNN by optimizing the architecture and the hyperparameters on both the WEAVE mocks and VIPERS observed spectra.
Table~\ref{tab:cnn_architecture} in Appendix~\ref{app:arc} summarizes the updated CNN architectures for both WEAVE and VIPERS data.

\subsection{U-Net}

The U-Net is a deep learning model initially designed for biomedical image segmentation \citep{ronneberger2015unet, zhang2018unet, li2024lensing}, which is effective and versatile. 
Unlike traditional CNNs that struggle with pixel-level predictions, U-Net excels by leveraging a unique encoder-decoder structure that captures context and fine details simultaneously.
The architecture's encoder reduces the input image's spatial dimensions through a series of convolutional and max-pooling layers. 
In this process, the model learns complex features and hierarchical representations. 
Instead, the decoder reconstructs the spatial dimensions using transposed convolutions and concatenations, combining them with the encoder's corresponding feature maps.
A key aspect of U-Net is its skip connections, which directly link the corresponding encoder and decoder layers. 
These connections ensure that high-resolution features are preserved and propagated through the network, enhancing the model’s ability to produce detailed and accurate segmentations.
Table~\ref{tab:unet}, in Appendix~\ref{app:arc}, reports the specific characteristics of the architectures we developed for the different data used in this work.


\section{Application to the spectra}\label{sec:application}

In this section, we describe the application of the NNs to the quasars (Sect.~\ref{subsec:app_qso}) and their generalization to galaxies (Sect.~\ref{subsec:app_gal}). In both cases, we start our analysis by considering the performance of the autoencoders, moving next to the analysis of the CNNs and U-Nets. 

\subsection{Application to the quasar spectra}\label{subsec:app_qso}

We give as input to the two autoencoders (the original IQNet and our bespoke model) the red part of the quasar spectra (${1216\, \text{\AA} < \lambda < 2000\, \text{\AA}}$ in the rest-frame) and predict the continuum in the entire spectrum range (${1020\, \text{\AA} < \lambda < 2000\, \text{\AA}}$ in the rest-frame), as was done in \citet{liu2021quasar} and  \citet{turner2024lycan}.
The red part of the quasar spectra is generally more regular than the blue side at these high redshifts because the red part is less contaminated by absorption lines that give rise to a thick forest that blankets the quasar continuum.
However, as an additional test, we examine whether passing the entire spectrum during training would improve the goodness of fit in the Ly$\alpha$ forest region.

The performances of autoencoders are shown in Fig.~\ref{fig:dist_affe_weave} (top panels). Each histogram reports the AFFE metric for the quasar continuum fit using the original iQNet and our bespoke autoencoder, training both on the full spectrum and on the red part.
In the three panels, we show instead the AFFE for the whole spectrum and in the red ($\lambda >= 1216~\text{\AA}$) and blue ($\lambda < 1216~\text{\AA}$) parts, separately.
Although the performance of iQNet is already satisfactory ($<1$ per cent deviations), our tailored architectures find a median AFFE consistently lower than iQNet's. The iQNet architecture also shows a more widespread distribution than our tweaked architectures.
Figure~\ref{fig:fit_weave} shows an example fit of a quasar spectrum.
Appendix~\ref{app:fit} contains more examples of fit performed with the autoencoders.

\begin{figure*}
    \centering
    \resizebox{\hsize}{!}{\includegraphics{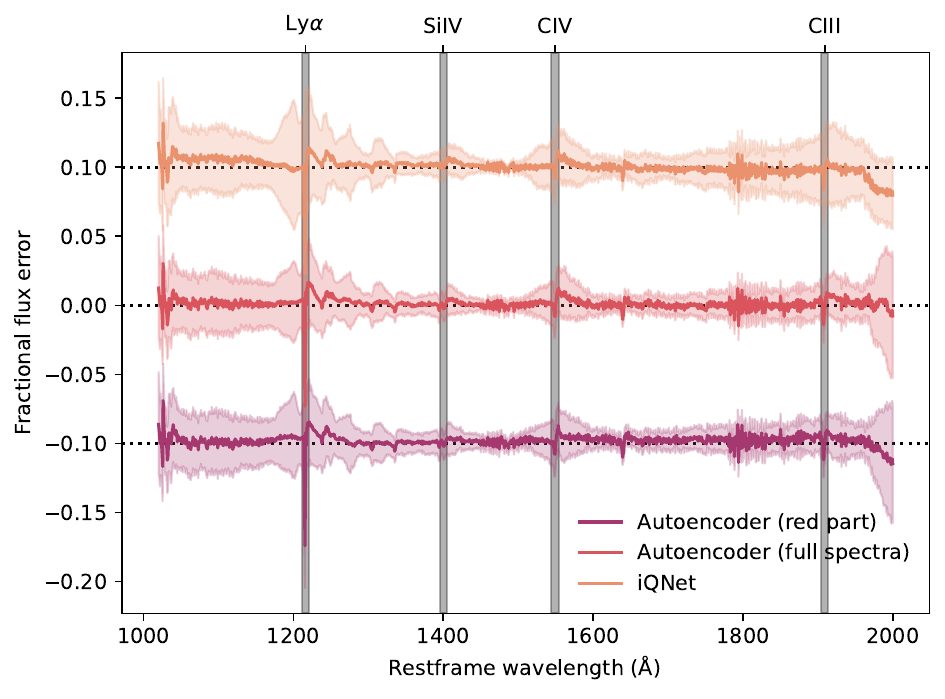}\includegraphics{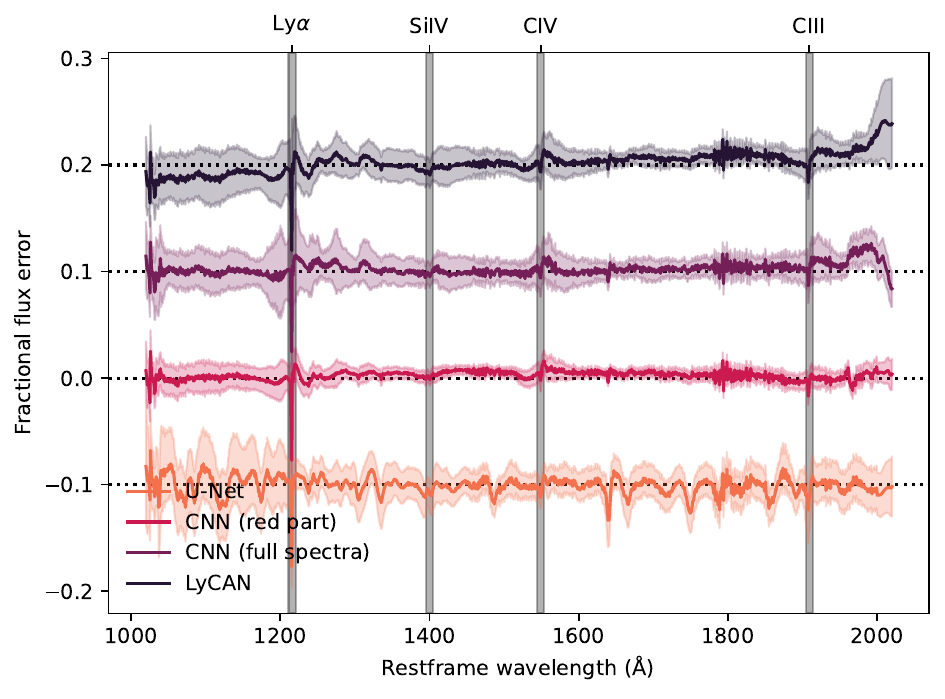}}
    \caption{Performance of NN predictions with wavelength: median fractional flux error (FFE) as a function of the wavelength for the autoencoders (left panel) and CNN (right panel). The shaded area shows the $16^{\rm th}$ and 84$^{\rm th}$ percentiles of the distribution of the whole sample. The vertical shift is introduced to improve visualization. The black dotted lines show the zero point of each run. The shaded black vertical lines show the positions of the main emission lines in the wavelength range.}
    \label{fig:ratio_weave}
\end{figure*}

The AFFE provides a global metric for the performance of the architectures. Instead, to examine whether any systematic bias arises as a function of wavelength, we study the performance on selected parts of the spectrum considering the FFE as a function of wavelength (Fig.~\ref{fig:ratio_weave}).
We found an FFE almost constant in all cases.
The overall small scatter, especially for the case of the tailored autoencoder, makes the application of this method reasonable to most problems where low-amplitude biases are not critical. However, once the FFE is looked at over the whole wavelength range, iQNet shows a bias (negative slope), making the bespoke architecture preferable given the fluxes are consistent with zero over an extensive range of wavelengths for $\lambda \geq 1216$ \AA.
The negative slope persists even though the training and test samples are shuffled. This suggests that iQNet requires a larger sample to reduce the bias.
Moreover, we did not find strong biases in the performances of the autoencoders with the S/N of the spectra or the redshift of the source (see Appendix~\ref{app:bias} for the details).

We next move to the analysis of the performance of the NNs, testing against each other the U-Net and CNN architecture, also in comparison with the results of the LyCAN architecture \citep{turner2024lycan}.
Examining the AFFE distribution (Fig.~\ref{fig:dist_affe_weave}), the U-Net performs slightly worse than the CNN and LyCAN, especially in the region of the Ly$\alpha$ forest. However, the performance over the entire spectrum is comparable among U-Net and LyCAN with a difference of a maximum of $0.2$ percent while the CNNs have performances equal to the autoencoders (top panels of Fig.~\ref{fig:dist_affe_weave}). 
Figure~\ref{fig:fit_weave} shows an example fit of a quasar spectrum. More examples of fits performed with these architectures are shown in Appendix~\ref{app:fit}.

We again examined possible biases in the continuum reconstruction in particular parts of the spectrum and studied the FFE as a function of the wavelength (Fig.~\ref{fig:ratio_weave}).
We found an FFE mainly constant in these cases. 
The LyCAN shows a small bias, with a positive slope.
The U-Net appears instead less prone to biases locally, although there are more fluctuations in the FFE compared to the other CNNs. As already noted when studying the AFFE, the autoencoder also performs better in this metric.
Checking for biases of the performance, we found a strong bias of the AFFE for the U-Net with the S/N, where better quality spectra have better-predicted continua.

We conclude that both autoencoders and CNNs perform well in reconstructing the quasar continuum, with the former showing better precision, accuracy, and stability. 
The reduced computational time and effort of the autoencoders ($\approx 10$ times lower than CNNs and U-Nets) is another aspect in favor of this class of NNs.

\section{Astrophysical tests and generalization}\label{sec:phys_test}

Having compared the performance of various NNs on the datasets that have been used for training, we explore further the ability of these architectures to be deployed on similar yet non-identical datasets without any training or new optimization. This section addresses how well these methods can be generalized to new datasets. Specifically, we consider a quasar sample from DESI onto which we deploy the NNs. Following continuum normalization using the best-performing NNs, we extract physical information on the mean transmitted flux for the Ly$\alpha$ forest in quasars which we compare to the measurements from the literature.

\begin{figure*}
    \centering
    \resizebox{\hsize}{!}{\includegraphics{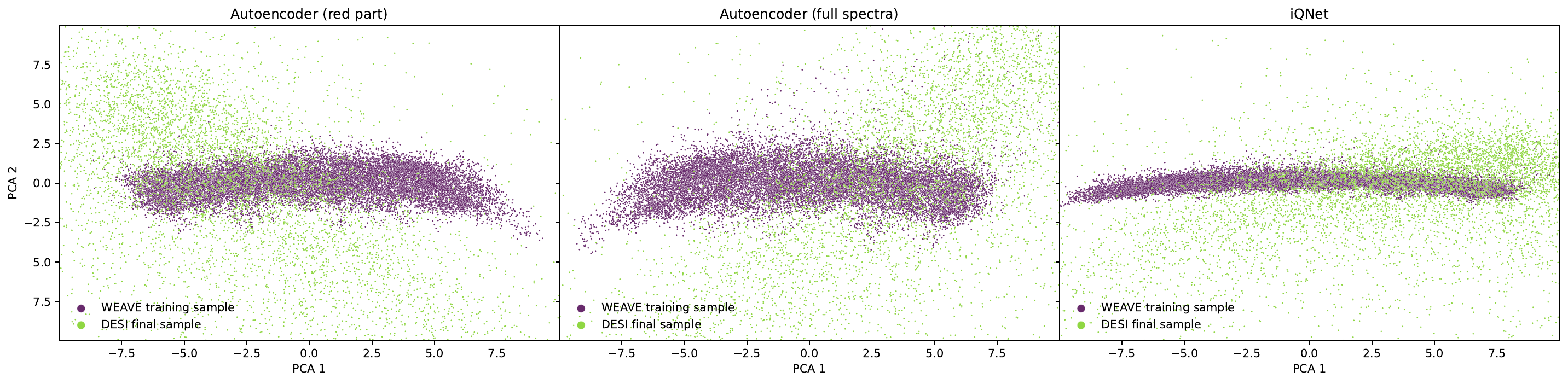}}
    \resizebox{\hsize}{!}{\includegraphics{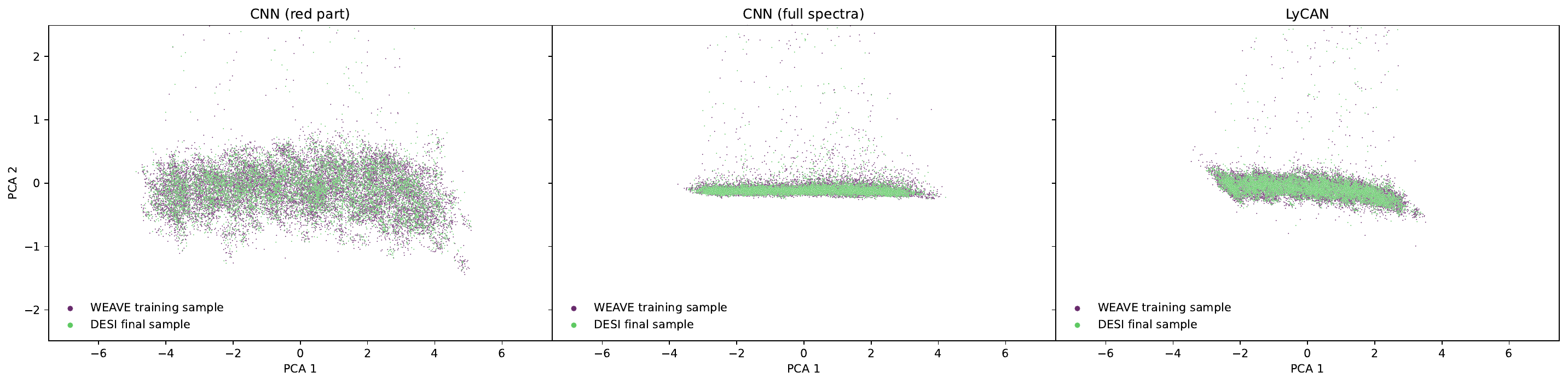}}
    \caption{Distribution of the spectra in the latent spaces of the autoencoders (upper row) and CNNs (bottom row) for the WEAVE training sample (purple) and the DESI final sample (green).}
    \label{fig:latent}
\end{figure*}



\subsection{Testing covariate shift}

When considering the application of NNs trained on different datasets, we should closely examine the impact of the covariate shift problem \citep{sanchez2014desi, hoyle2015sdss_dataaugment, rau2015photoz, zitlau2016sdss}.
A covariate shift occurs when the input distribution changes between training and testing samples, causing the model to perform poorly on new data. This shift can lead to mismatches between learned representations and actual test distributions.

We consider the minimal representation of the input data \citep[training sample of WEAVE mocks and DESI data,][]{ambrosch2023chemgaia, ginolfi2025moons, belfiore2025dann} from each NN, corresponding to the central layer for the autoencoders and the dense layer with the lowest dimension for the CNNs. Then, we perform PCA on WEAVE data, and we apply the same transformation to both samples. The PCA representation of the data is then plotted in a 2D space (the first component has $\sim 90\%$ of variance for all cases, see Fig.~\ref{fig:latent}) for all NNs. We observe that the DESI sample explores a wider portion of the latent space for the autoencoders with respect to the WEAVE training sample.  CNNs, instead, can map both samples in a more similar latent space.
We choose to use the linear transformation given by the PCA because it allows us to be consistent in the comparison of the two samples. Non-linear transformations, such as uniform manifold approximation and projection (UMAP) or t-distributed stochastic neighbor embedding (t-SNE), would break the consistency between samples by their nature, leading to a more difficult comparison of the latent spaces.

To study the effects of training in a subregion of the latent space and then applying the model to a sample that occupies a separate subregion, we split the 2D representation of the training sample into two parts at $\mathrm{PCA}=0$: vertically (left and right) and horizontally (bottom and top).
By doing this, we reduce the sample size for the training by half and predict the continuum on spectra that are not close to those used for the training. Splitting both vertically and horizontally is useful for breaking the symmetries in the distributions of the spectra in the latent space.

After splitting the 2D representation of the training sample, we retrain each NN again on the left and test it on the right, and vice versa.
Similarly, we repeat the same procedure for the horizontal splitting. 
Despite the presence of predominating vertical symmetry, we do not observe strong biases in the distributions, with most of the AFFE values below $\approx 0.02-0.03$ for the autoencoders. However, in some cases, we observe an excess at the tail of the AFFE distributions for high values that remain nevertheless below $\approx 0.05-0.06$ while the CNN vertically splitted reach values of $0.10$.
Even in the most pessimistic case, where the NNs have been trained only on half the latent space, the performances decrease by a factor $\approx 2$ or $3$ for some cases with the predicted continua, with an AFFE associated with $\approx 2.5$ percent when an excess in the distribution is observed.

From this exercise, we conclude that the NNs included in this work can also be successfully deployed on different datasets than the one employed during training.
As a last exercise, we apply the best-performing NNs to two astrophysical problems that rely on a precise and accurate reconstruction of the continuum level in quasars and galaxies, using the DESI EDR sample. 

We estimate the mean transmitted flux in the Ly$\alpha$ region (Sect.~\ref{subsec:optical}).
We performed the test with the continua estimated with our optimized autoencoder that consistently displays good performance metrics. 

\begin{figure}
    \centering
    \resizebox{\hsize}{!}{\includegraphics{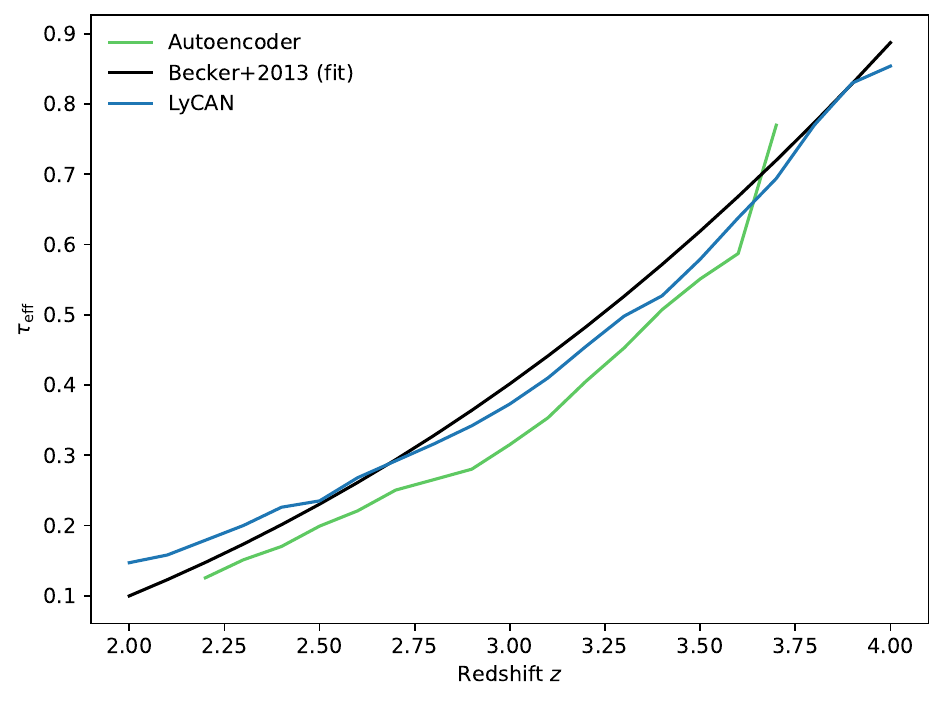}}
    \caption{Redshift evolution of the effective optical depth ($\tau_\mathrm{eff}$) observed in DESI EDR sample. We report the values obtained in this study with the continua estimated with the optimized autoencoder (green). The evolution found by \cite{becker2013tau} (black) and the results on the DESI observation done with LyCAN \citep{turner2024lycan} (blue) are also shown.}
    \label{fig:tau}
\end{figure}

\subsection{Optical depth of the Ly$\alpha$ forest}\label{subsec:optical}

For a more physical test of quasar continua, we estimated the mean transmitted flux, converted as optical depth $\tau$, in the region of the Ly$\alpha$ forest assuming the continua derived by the autoencoder.
The weights obtained during the training phase with the WEAVE mock spectra are directly applied to the DESI EDR catalog without any new training to estimate the continua of the spectra.
We then take the portion of the spectra in the rest-frame wavelength range $1\,070~\text{\AA}$--$1\,160~\text{\AA}$ to avoid contamination from the Ly$\alpha$ and $\ion{O}{vi}$ \citep{kamble2020tau}.
At each pixel of the spectra in this range, we associated a redshift $z_\mathrm{Ly\alpha}$ of the absorber given by the equation:
\begin{equation}
    z_\mathrm{Ly\alpha} = \frac{\lambda \left(1 + z_\mathrm{QSO} \right)}{\lambda_\mathrm{Ly\alpha}} - 1.
\end{equation}
In each $z_\mathrm{Ly\alpha}$ bin (binwidth of $0.1$), we compute mean transmission defined as
\begin{equation}
     f \left( z_\mathrm{Ly\alpha} \right) = \frac{F_\mathrm{obs} \left( z_\mathrm{Ly\alpha} \right)}{F_\mathrm{cont} \left( z_\mathrm{Ly\alpha} \right)},
\end{equation}
where $F_\mathrm{obs}$ is the observed flux and $F_\mathrm{cont}$ is the estimated continuum at the corresponding $z_\mathrm{Ly\alpha}$.
Finally, we computed the median transmitted flux $\left< f  \right>$ and converted it to optical depth ${\tau = - \ln \left< f \right>}$.
We emphasize that this analysis aims to test whether we can recover previous estimates of this quantity and not present a novel measurement of the IGM mean optical depth. As such, we report the raw quantity derived as defined above, referring the readers to detailed literature on the subject for accurate measurements \citep{aguirre2002method, schaye2003metals, becker2013tau, turner2024lycan}.
We also do not apply any correction for optically thick absorbers \citep{becker2013tau} or metals \citep{schaye2003metals, kirkman2005igm, faucher2008igm}.

Figure~\ref{fig:tau} shows the evolution of our measurement of the optical depth.
Our simple approach can reproduce quite well the observed evolution in more bespoke studies \citep{becker2013tau, turner2024lycan}, which implies a good degree of prediction of the quasar continuum in unseen DESI data. For comparison, we reported in the same figure the tabulated values in \cite{turner2024lycan} (table~3, final values, bias-corrected measurements corrected for metal line absorption according to \citealt{schaye2003metals} and optically thick absorbers according to \citealt{becker2013tau}) and the best-fit relationship of \cite{becker2013tau}.
The autoencoder thus demonstrates an excellent degree of generalization, so that it can be trained on mock spectra and applied without further tuning to comparable but not identical data.

\begin{figure*}
    \centering
    \resizebox{\hsize}{!}{\includegraphics{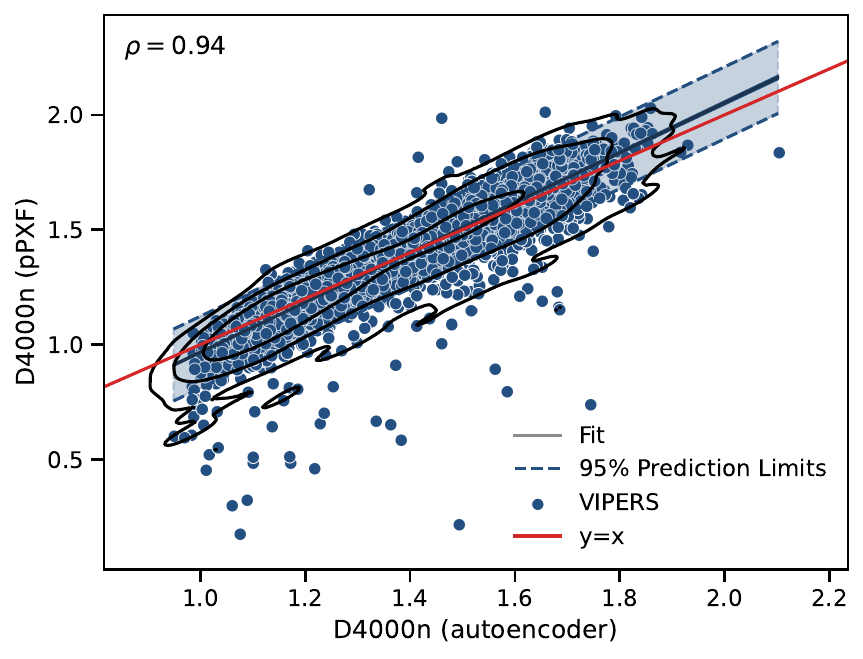}\includegraphics{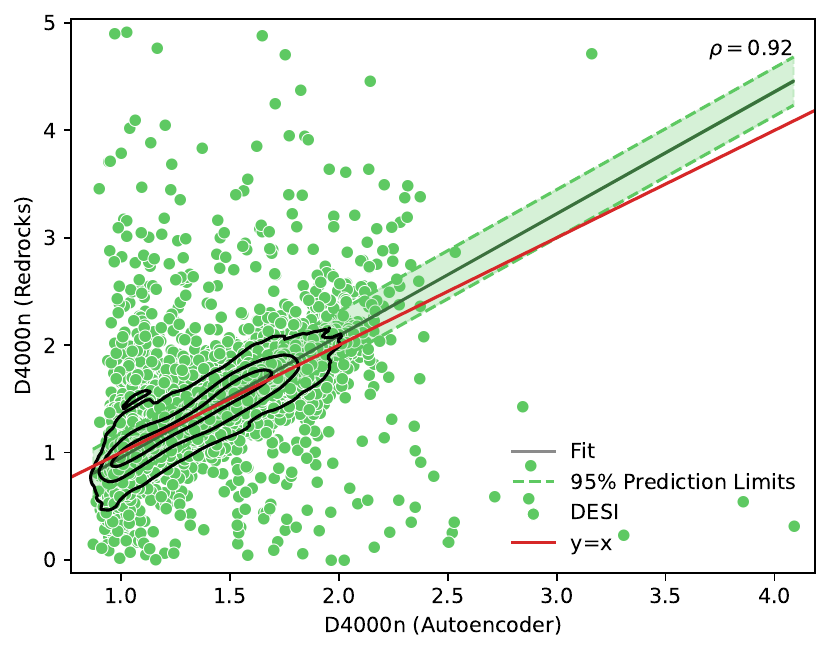}}
    \resizebox{\hsize}{!}{\includegraphics{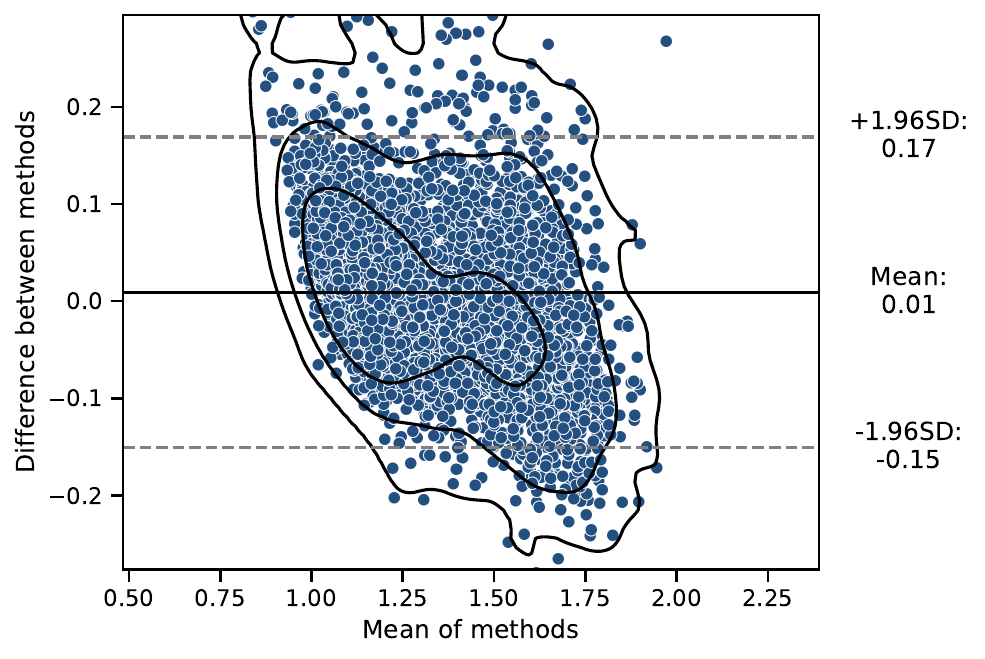}\includegraphics{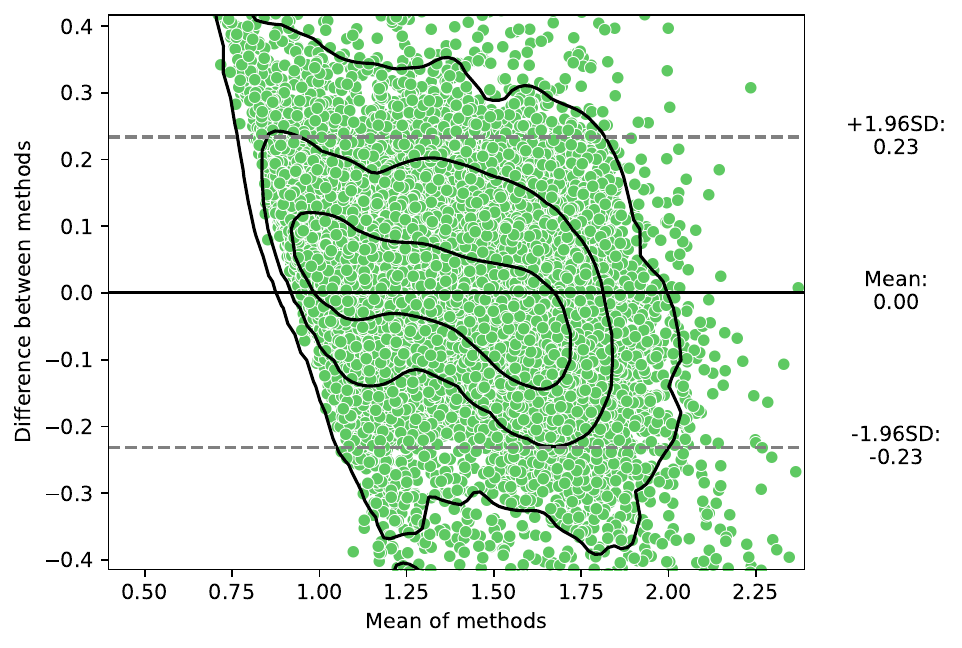}}
    \caption{Comparison of the D4000n values estimated with the NN in this work and with pPXF \citep{pistis2024cross} (left panel) for VIPERS (in blue) and with Redrock (right panel) for DESI (in green). The red solid line shows the line $y=x$, the light-shaded area limited by dashed lines shows $95\%$ prediction limits and the dark-shaded area shows the $95\%$ confidence limits. The density plots show the one-, two-, and three-sigma levels of the 2D distributions. Top panel: direct comparison of the two methods. Bottom panel: Bland–Altman plot.}
    \label{fig:d4000n}
\end{figure*}

\section{A note on computational efficiency}\label{sec:eff}

The above analysis demonstrates that the autoencoder provides a lightweight yet powerful alternative to more complex architectures when determining the continuum of quasars. 
An added benefit of this simplicity is that the autoencoder is much less computationally expensive than the U-Net and CNNs, by a factor of $\sim 20$ in computational time, resulting in a better choice when analyzing a large data sample is necessary.
The difference in performance of $0.2\%$ between the autoencoder and the CNN is negligible with respect to the difference in computational time, especially when optimization and training steps must be performed.

\begin{figure}
    \centering
    \resizebox{\hsize}{!}{\includegraphics{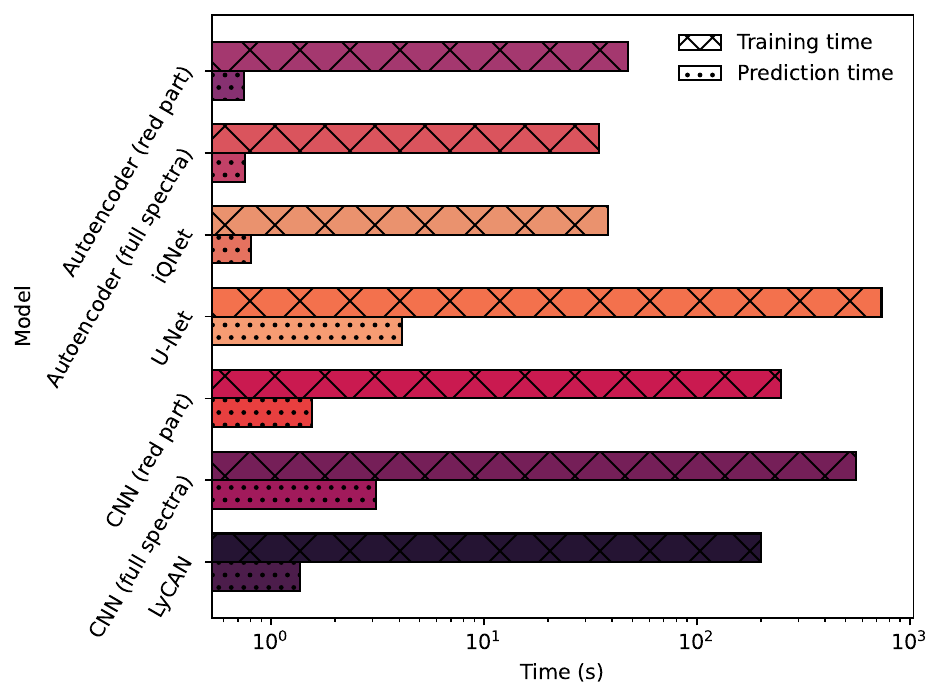}}
    \caption{Visualization of the training and prediction times for the autoencoders, U-Net, and CNNs.}
    \label{fig:times}
\end{figure}

Figure~\ref{fig:times} visualizes the training and prediction times for all NNs using three NVIDIA A2 GPUs. The training and prediction phases are performed using the TensorFlow mirrored strategy to distribute the work between the three GPUs.
For a fair comparison, the batch size is set to 256 spectra (limited by avoiding memory problems with the U-Net and CNNs) for all NNs, but, in principle, the batch size for autoencoders can be easily increased, reducing even further the computational time.
The main difference is found in the training time, where U-Net and CNN have an order of magnitude longer time. 
This is even more visible during the optimization of the hyperparameters, when the autoencoders can perform 100 trials in a fraction of the time needed for 50 trials for the U-Net and CNNs.
Moreover, the computational time does not increase linearly with the sample size.
Assuming a sample size of $1\,000\,000$ spectra to fit, an autoencoder would spend about $3$ minutes, while a CNN would spend about $40$ minutes.

\begin{figure}
    \centering
    \resizebox{\hsize}{!}{\includegraphics{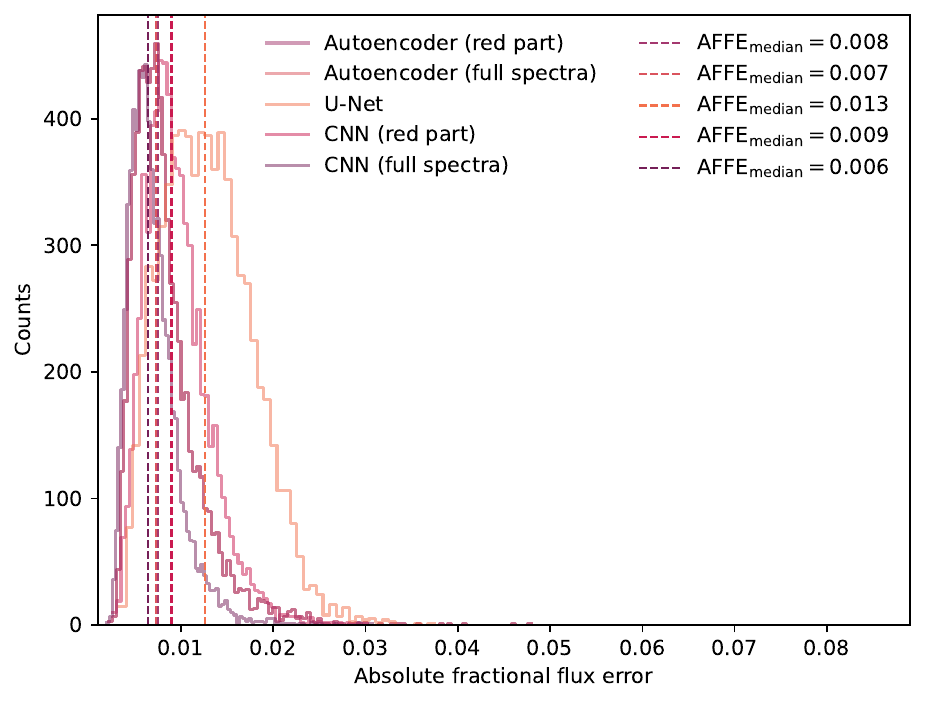}}
    \caption{Performance of NN predictions: histograms of the absolute fractional flux error (AFFE) for NNs applied to the WEAVE mock catalog for the novel architectures used in this work.}
    \label{fig:affe_best}
\end{figure}

\section{Test on galaxies}\label{sec:gal}

Having investigated the performance of each NN on quasars, we move to the questions of whether these architectures can be generalized to different continuum shapes with minimal modfications (i.e., new optimization and training steps). Bespoke NNs tailored for the analysis of galaxy spectra can be found in the literature, e.g. SPENDER \citep{melchior2023spender, liang2023outdet, liang2023aegal}. Applications with a focus on galaxies should therefore turn to these dedicated networks. This section aims solely at testing the performance of NNs developed for quasars when applied (nearly) out-of-the-box on galaxy spectra.

We thus focus our attention on the performance of each architecture in predicting the shape of galaxy spectra from the VIPERS survey. Compared to a mostly featureless power-law continuum plus emission lines, galaxies present a richer set of features encoded in their stellar continuum. 
Without heavily restructuring the algorithms, the architectures used for the quasars undergo a new optimization step and re-training using a sub-sample of VIPERS spectra. In this way, we check the performance of the NNs with a limited tuning of the methodology. We give as input to the NNs the whole galaxy spectra (${3500\, \text{\AA} < \lambda < 5500\, \text{\AA}}$ in the rest-frame) and predicted the continuum in the same range.
We selected galaxies within a redshift range of $0.5 \leq z \leq 0.75$ to have the observations within the whole wavelength range.
No particular additional data selection (for example, a threshold on the $\mathrm{S/N}$ values of the spectra) was applied to the training sample.

\begin{figure}
    \centering
    \resizebox{\hsize}{!}{\includegraphics{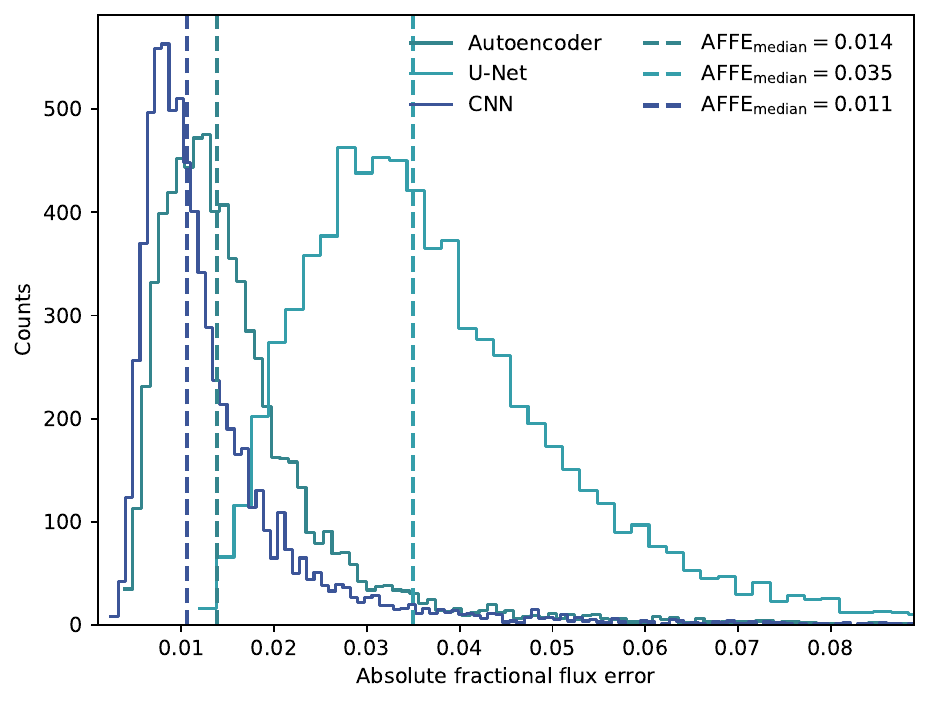}}
    \caption{Performance of NN predictions: histograms of the absolute fractional flux error (AFFE) for NNs applied to VIPERS galaxies for the novel architectures used in this work.}
    \label{fig:affe_vipers}
\end{figure}

Figure~\ref{fig:affe_vipers} shows the AFFE distributions for the NNs applied to galaxy spectra.
As in the quasar case, the autoencoder and CNN perform better than the U-Net for the fit of the continua, with an AFFE of $\approx 1$ per cent with respect to the continua estimated using traditional methodology such as pPXF \citep{cappellari2004fit, cappellari2017improving, cappellari2022full}.
The same methodology used for the quasars can be applied to galaxies with an increased median AFFE of $\approx 0.5$ per cent.

Also for galaxies, we tested the ability of our autoencoder to make reliable predictions on unseen DESI galaxy spectra. Fig.~\ref{fig:d4000n} shows the comparison of the values for the D4000n break calculated over the continua predicted by the autoencoder and the continua estimated using pPXF for VIPERS and Redrock for DESI.
We stress again that the use of the continua made with Redrock is dictated by its availaility in the DESI EDR catalogs, and bespoke algorithms within the full DESI pipeline will outperform this algorithm in continuum determination. Still its use is sufficient to reach our main conclusion on the generalization capability of the NNs explored in this study.

We found a value of the Pearson correlation coefficient of $0.94$ and $0.92$ for VIPERS and DESI, respectively, between the D4000n break calculated over the continua predicted by the NN and the reference continua (estimated with pPXF for VIPERS and Redrock model for DESI).
In the Bland-Altman plot \citep{bland1986statistical}, both data sets show a small shift (bias) of the means from 0, however, for both VIPERS and DESI cases, the data show a hint of a negative slope.
For these datasets, the bias is proportional to the measurements.
We thus conclude that also for the galaxy spectra, the autoencoder generalizes sufficiently well to unseen data, and it is suitable for astrophysical problems.


\section{Summary}\label{sec:sum}

This study focus on a comparative analysis of the performance of various NNs to predict the continuum of quasars.
Specifically, we compared the performance of an autoencoder, a CNN, and a U-Net with two published architectures (iQNet from \citealt{liu2021quasar} and LyCAN from \citealt{turner2024lycan}). 
Initial comparison was performed on a large catalog of mock spectra for the WEAVE survey \citep{jin2024weave}.
Using the DESI EDR quasar catalog \citep{desi2022overview, desi2023edr}, 
we further tested the ability of these architectures to generalize to unseen datasets without the need of retraining. 
Moreover, with minor optimizations, we studied to what extent NNs tailored to handle quasar spectra can also be deployed on more complex spectral shapes, such as galaxies. For this task, we rely on galaxies observed with the VIPERS survey \citep{garilli2014vimos, scodeggio2018vimos, vergani2018nev, vietri2022agn, figueira2022sfr, pistis2022bias, pistis2024cross} and on the DESI EDR galaxy catalogs.
Fig.~\ref{fig:affe_best} summarizes the performances of the various architectures applied to the quasar problem, using the AFFE distributions as a metric.

The optimized autoencoder developed in this work improves the width of the AFFE distribution when applied to $z \gtrsim 2$ quasars with respect to the original iQNet by \cite{liu2021quasar} and reaches better performances than those of \cite{turner2024lycan} with LyCAN (see Sect.~\ref{sec:application}).
The autoencoder also performs well with galaxy spectra after a retraining and optimization step, but without restructuring the architecture (see Fig.~\ref{fig:affe_vipers} for both passive and active sources). While custom NNs tailored to the galaxy problem can be found in the literature, this test shows a good versability of the architectures explored in this work to a variety of continuum shapes.

Finally, we directly applied the autoencoder to the DESI EDR data (see Sect.~\ref{sec:phys_test}) without retraining, for both the quasar and the galaxy case.
With the predicted continua, we estimated the evolution of the mean transmitted flux in the Ly$\alpha$ forest region converted to optical depth.
We found an evolution in agreement with previous studies \citep{becker2013tau, turner2024lycan} indicative of the possibility of using these NNs in previously unseen datatsets from different surveys.
Likewise, for galaxies, we estimated the D4000n break with the continua estimated with the autoencoder for both VIPERS and DESI galaxy catalogs and compared them with the values with the continua estimated with pPXF for VIPERS and the model, done with Redrock, released with the DESI EDR catalog.
In both cases, the linear fit is close to the $y=x$ line, with a deviation around $\mathrm{D}4000\mathrm{n}= 0.2$.

The main conclusions can be summarized as follows:
\begin{itemize}
    \item the autoencoder performs as well as more complex architectures for much lower computational costs;
    \item the autoencoder performs well with both quasar and galaxy spectra;
    \item the direct application to DESI EDR data confirms the generalization of the method to unseen datasets which are similar but not identical to the training data;
    \item all the architecture tested in this work show a small bias of the FFE with the wavelength related to the magnitude selection of the sample (see Appendix~\ref{app:ffe_bias}), and high-precision tasks should identify suitable metrics for assessing the impact of these offsets.
\end{itemize}

A simple autoencoder proves useful for reaching good precision (median error of $\approx 1$ per cent) in the continuum predictions while limiting computational resources in analyzing large data samples. This NN can become a versitile tool for studies about absorbers in the quasar spectrum, for example tracking the evolution of strong absorbers in the CGM, such as \ion{C}{iv} \citep[for example,][]{cooksey2013civ, davies2023civcat, davies2023civ}, \ion{Si}{iv} \citep[for example,][]{dodorico2022siiv}, and \ion{Mg}{ii} \citep[for example,][]{matejek2012mgii, chen2017mgii, zou2021mgii} or the auto- and cross-correlation of Ly$\alpha$ absorbers \citep[for example,][]{dumas2020bao}.




\begin{acknowledgements}


We thank the referee for insightful comments that have improved the content and presentation of this work. 

F.~Pistis, M.~Fumagalli, and M.~Fossati have been supported by the European Union -- Next Generation EU, Mission 4, Component 1 CUP H53D23011030001.

W.~J.~Pearson has been supported by the Polish National Science Center project UMO-2023/51/D/ST9/00147.

This paper uses data from the VIMOS Public Extragalactic Redshift Survey (VIPERS). VIPERS has been performed using the ESO Very Large Telescope, under the ``Large Programme'' 182.A-0886. The participating institutions and funding agencies are listed at \url{http://vipers.inaf.it}.

This research used data obtained with the Dark Energy Spectroscopic Instrument (DESI). DESI construction and operations are managed by the Lawrence Berkeley National Laboratory. This material is based upon work supported by the U.S. Department of Energy, Office of Science, Office of High-Energy Physics, under Contract No. DE–AC02–05CH11231, and by the National Energy Research Scientific Computing Center, a DOE Office of Science User Facility under the same contract. Additional support for DESI was provided by the U.S. National Science Foundation (NSF), Division of Astronomical Sciences under Contract No. AST-0950945 to the NSF’s National Optical-Infrared Astronomy Research Laboratory; the Science and Technology Facilities Council of the United Kingdom; the Gordon and Betty Moore Foundation; the Heising-Simons Foundation; the French Alternative Energies and Atomic Energy Commission (CEA); the National Council of Science and Technology of Mexico (CONACYT); the Ministry of Science and Innovation of Spain (MICINN), and by the DESI Member Institutions: www.desi.lbl.gov/collaborating-institutions. The DESI collaboration is honored to be permitted to conduct scientific research on Iolkam Du’ag (Kitt Peak), a mountain with particular significance to the Tohono O’odham Nation. Any opinions, findings, and conclusions or recommendations expressed in this material are those of the author(s) and do not necessarily reflect the views of the U.S. National Science Foundation, the U.S. Department of Energy, or any of the listed funding agencies.
\end{acknowledgements}

\bibliographystyle{aa}
\bibliography{Bibliography}

\begin{appendix}

\section{Bias on the fractional flux error}\label{app:ffe_bias}

\begin{figure*}[!h]
    \centering
    \resizebox{\hsize}{!}{\includegraphics{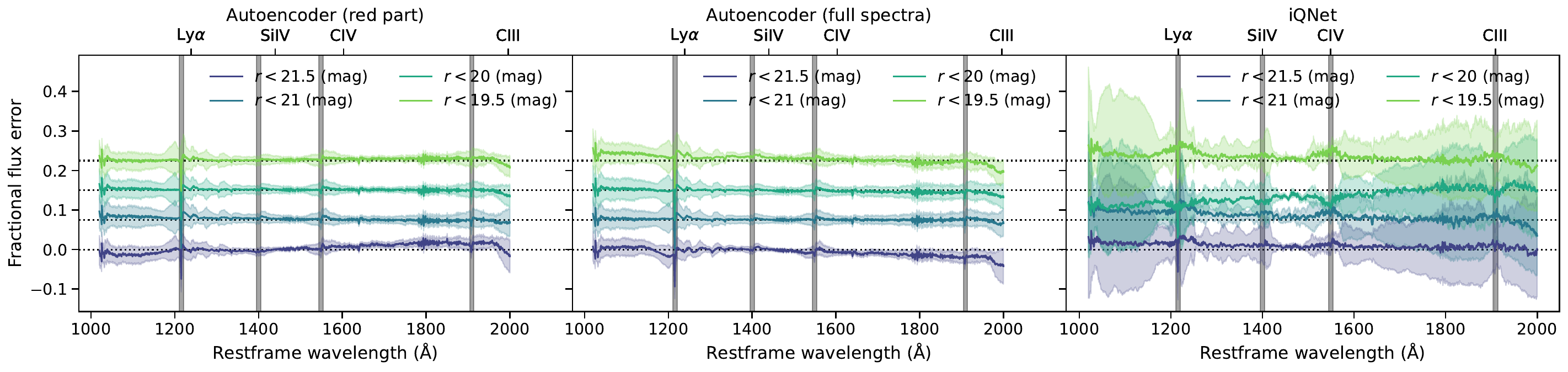}}
    \resizebox{\hsize}{!}{\includegraphics{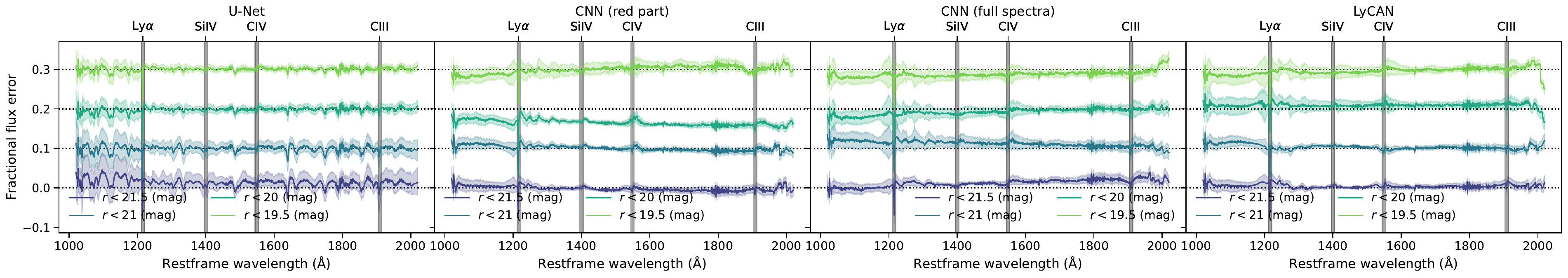}}
    \caption{Fractional flux error as functional of the rest-frame wavelength for the autoencoders (top panels) and to the CNNs (bottom panels) for different selection on the $r$-band magnitude.}
    \label{fig:auto_bias}
\end{figure*}

Fig.~\ref{fig:auto_bias} (top panel) shows the FFE as a function of the rest-frame wavelength for the cases of passing the red part of the spectra or the whole spectra as input to the autoencoder and iQNet.
Different colors correspond to different $r$-band magnitude cuts of the sample.
In all cases, using less restrictive cuts improves the overall performance of the autoencoders.
iQNet seems to be more sensitive to magnitude selection.
We have applied a cut at $r < 21$ to minimize the bias while maximizing the sample statistics through our analysis.

Fig.~\ref{fig:auto_bias} (bottom panel) shows the FFE as a function of the res-frame wavelength for the case of U-Net, optimized CNNs and LyCAN.
Only the restrictive cut, such as $r < 19.5$, reduces CNN performance.
Performance in the Ly-$\alpha$ forest region improves with less restrictive cuts.
LyCAN is less sensitive to the magnitude bias, except the Ly-$\alpha$ forest region

\FloatBarrier

\section{Effects of S/N selection}\label{app:snr}

\begin{figure}
    \centering
    \resizebox{\hsize}{!}{\includegraphics{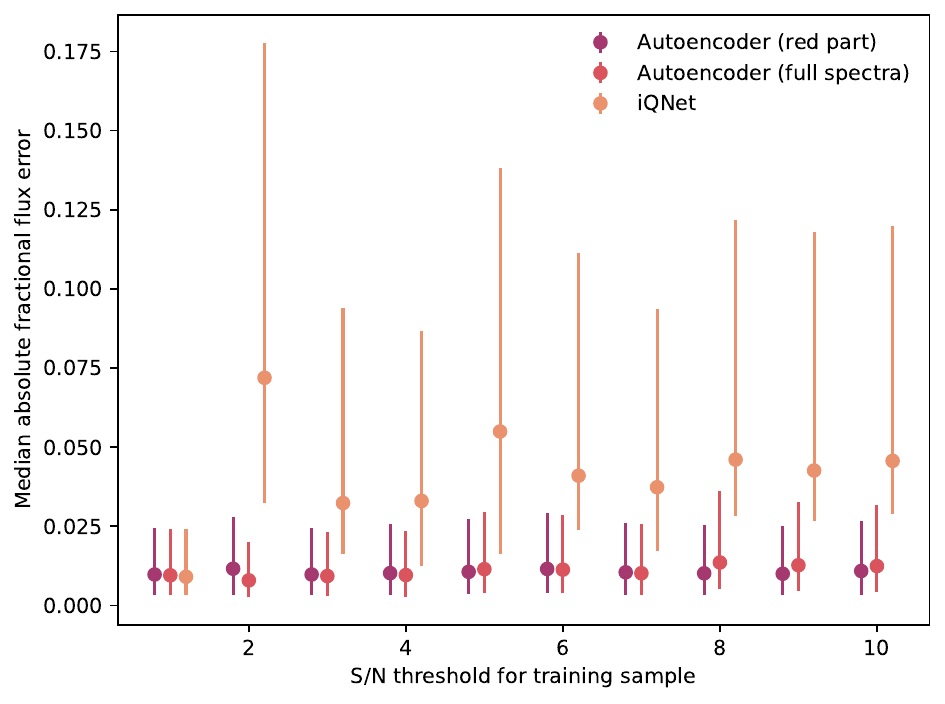}}
    \resizebox{\hsize}{!}{\includegraphics{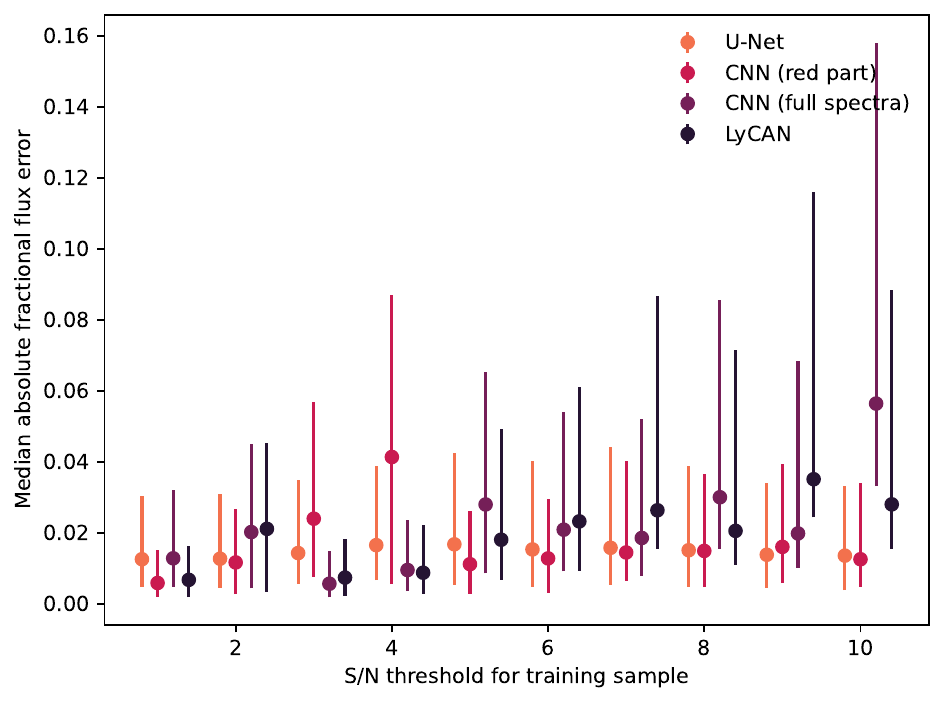}}
    \caption{Median AFFE as a function of the selection S/N-threshold for the autoencoders (top) and the CNNs and U-Net (bottom panel) for quasars. The error bars show the 16$^{\rm th}$ and 84$^{\rm th}$ percentiles of the AFFE distribution in each run. The horizontal shift between points is introduced to improve the visualization.}
    \label{fig:affe_vs_snthresh_weave}
\end{figure}

A relevant question in the training step is what $\mathrm{S/N}$ threshold should be imposed. Specifically, we want to understand whether the algorithm performs better when seeing high-quality data or if we should feed as much information as possible, presenting low-S/N spectra in the training step. 
To this end, before creating our training and test samples, we removed sources with a median $\mathrm{S/N}$ value of the spectra lower than a given threshold. These sources are then added directly to the test sample to check the performance in generalizing low-$\mathrm{S/N}$ data starting from high-$\mathrm{S/N}$ data.
Fig.~\ref{fig:affe_vs_snthresh_weave} (top panel) shows the results of this analysis.
For iQNet, we see hints of a break at $\mathrm{S/N}>2$, though when considering error bars, the median AFE appears to be roughly constant.
We conclude that the autoencoders can generalize the fit of high-S/N spectra for low-S/N spectra.
The iQNet can learn better from a larger dataset, including spectra at low S/N.
The analyses in this study are done without selecting the samples on the $\mathrm{S/N}$.

We follow the same approach for the CNNs.
Fig.~\ref{fig:affe_vs_snthresh_weave} (bottom panel) shows the results of this analysis.
For the U-Net and CNNs, the median AFFE and its scatter increase with the $\mathrm{S/N}$ threshold. The deviation is more evident for the CNN trained on the whole spectra and LyCAN. Thus, for all architectures, we conclude that training in larger, more comprehensive datasets is more useful than training in high-quality samples.

\FloatBarrier

\section{Model biases}\label{app:bias}

\begin{figure*}
    \centering
    \resizebox{.96\hsize}{!}{\includegraphics{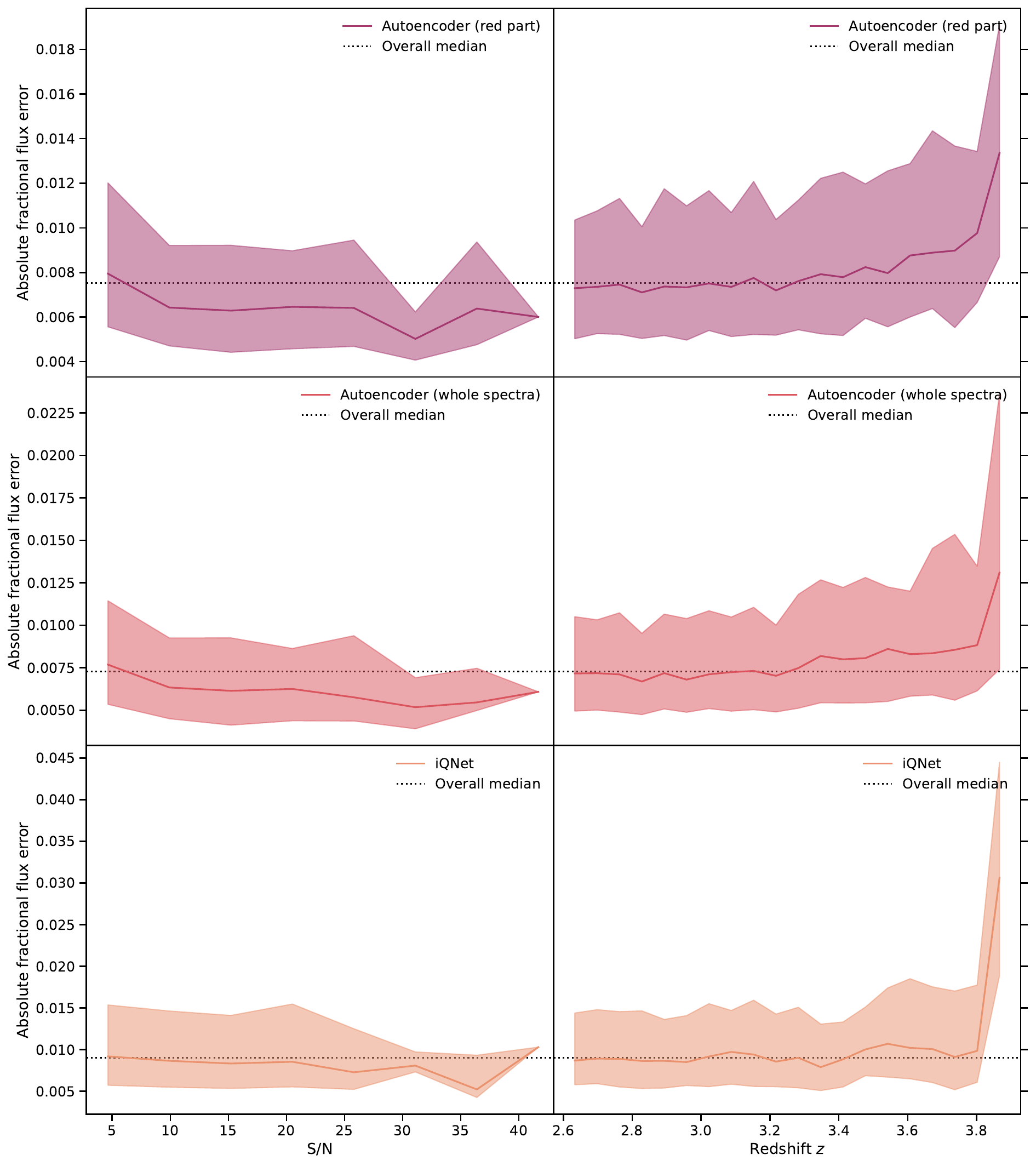}\includegraphics{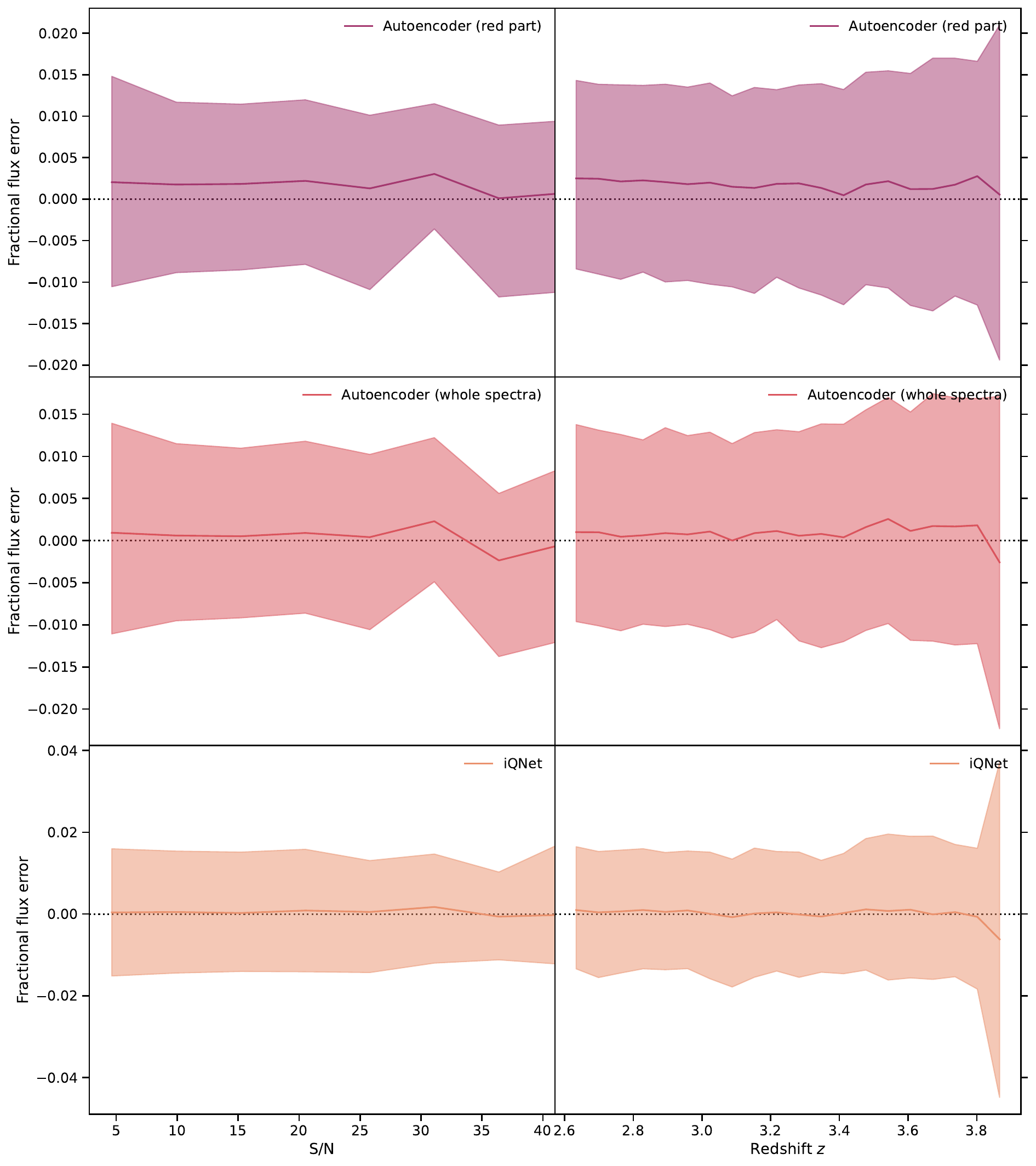}}
     \resizebox{.96\hsize}{!}{\includegraphics{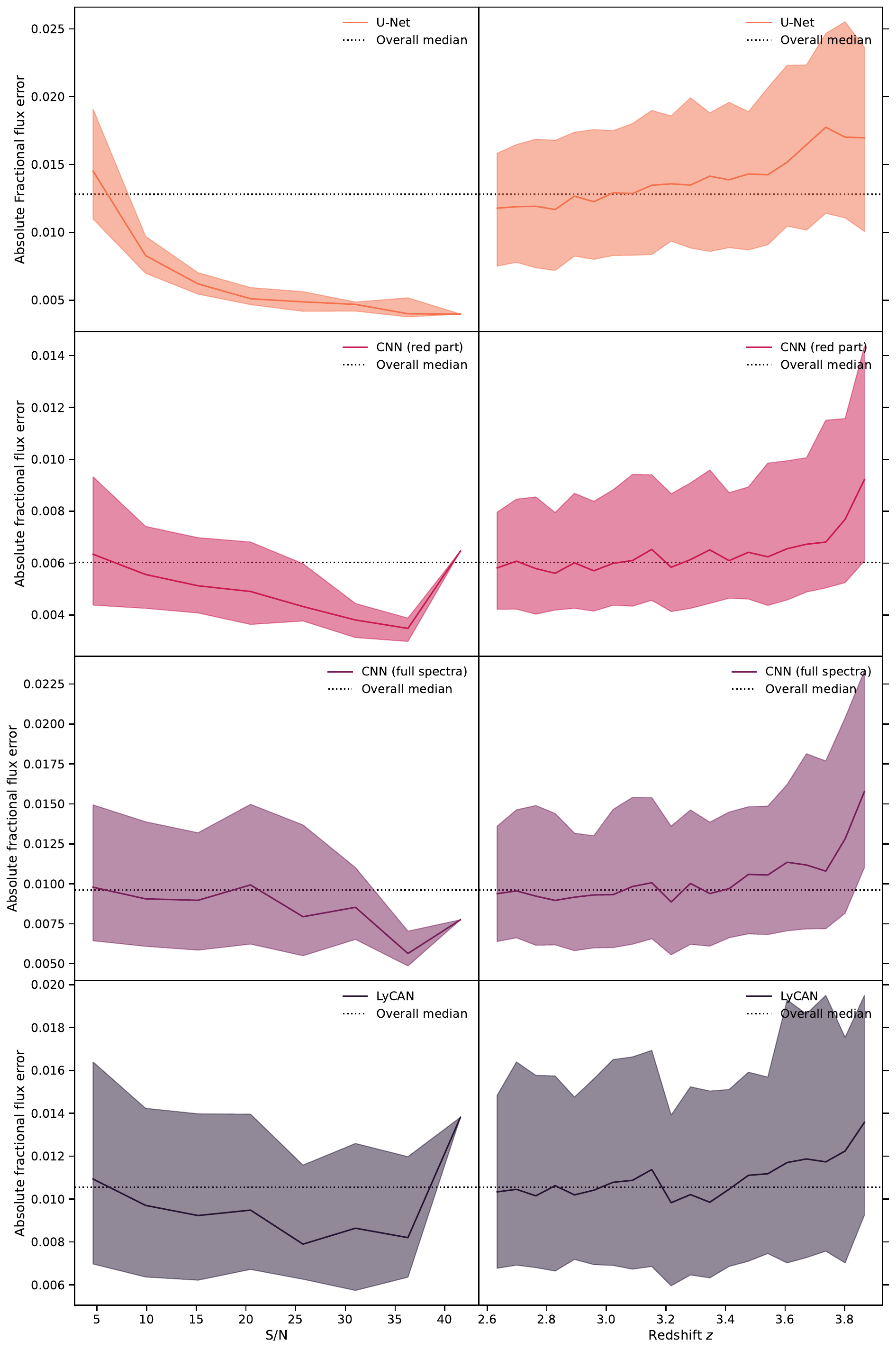}\includegraphics{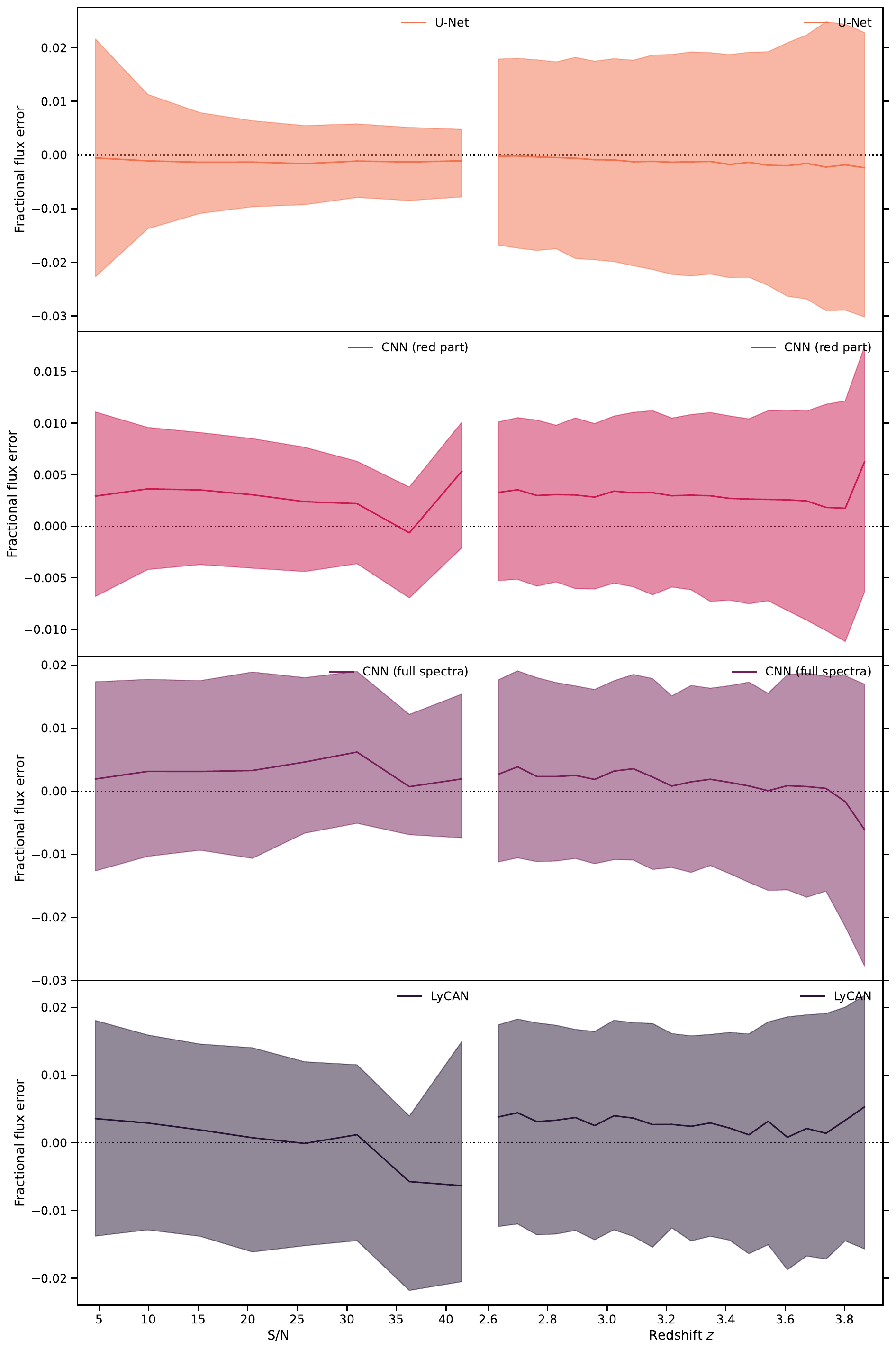}}
    \caption{Systematic biases of autoencoders (top row) and CNN plus U-Net (bottom row) for galaxies. The left (right) column shows the AFFE (FFE) as functions of $\mathrm{S/N}$ and $z$. The dotted lines show the median AFFE values and the zero-point for the FFE. The shaded region shows the 16$^{\rm th}$ and 84$^{\rm th}$ percentiles.}
    \label{fig:affe_vs_properties_weave}
\end{figure*}

As a final performance metric, we consider the dependence of the fit quality as a function of the $\mathrm{S/N}$ of each spectrum and redshift of each quasar. Fig.~\ref{fig:affe_vs_properties_weave} shows the median relation between AFFE (left panel) and FFE (right panel) versus $\mathrm{S/N}$ and redshift. For the AFFE, the median relation is almost constant in the whole $\mathrm{S/N}$ and redshift ranges, with a rise only for $z > 3.8$. The shift from the median in the last redshift bin is primarily due to the reduced statistic in that bin. The iQNet architecture shows a more widespread scatter around the full sample median. The situation is similar for the FFE, with the most shifted bin being the last redshift bin and the overall median relation constant. The iQNet architecture again shows a more widespread scatter.

As for the autoencoders, we checked the dependence of the quality of the fit on the $\mathrm{S/N}$ and redshift of each source also for the CNNs (Fig.~\ref{fig:affe_vs_properties_weave}).
U-Net, the CNN trained on the red part only, and LyCAN show a decreasing AFFE with the $\mathrm{S/N}$, a trend that is more evident in the case of the U-Net where the median relation drops below the full sample median.
The AFFE, instead, increases with the redshift, monotonically for the U-Net, especially above $z=3.5$, while it is generally constant for the other architectures, except for the highest redshift. Regarding the FFE, all architectures show a systematic shift in both $\mathrm{S/N}$ and redshift.
Only the U-Net underestimates the continuum at every $\mathrm{S/N}$ and redshift. In contrast, CNNs and LyCAN tend to overestimate the continuum.

\FloatBarrier

\section{Summary of architectures}\label{app:arc}

Tables~\ref{tab:iqnet}, \ref{tab:cnn_architecture}, and \ref{tab:unet} summarizes the optimized architectures used in this work.
The sizes of each layer and additional information, such as activation functions and kernel sizes (for convolutional layers), are explicitly reported in the tables.

\begin{table}[!h]
\caption{Architecture of the optimized autoencoders applied to WEAVE (red and full spectra) and VIPERS data.}
\label{tab:iqnet}
\centering
\begin{tabular*}{\columnwidth}{l @{\extracolsep{\fill}} l l l}  
\hline\hline                        
Model & Type & Size & Additional Info \\\hline
\noalign{\smallskip}
WEAVE  & Input & 3920 & \\
red part & Masking & 3920 & Mask value: 0 \\
 & Dense & 512 & Act.: ReLU \\
 & Dense & 128 & Act.: ReLU \\
 & Dense & 1024 & Act.: ELU \\ 
 & Dense & 4900 & \\
\hline
\noalign{\smallskip}
WEAVE  & Input & 4900 & \\
full spectra & Masking & 4900 & Mask value: 0 \\
 & Dense & 512 & Act.: ReLU \\
 & Dense & 128 & Act.: ReLU \\
 & Dense & 1024 & Act.: ELU \\
 & Dense & 4900 & \\
\hline
\noalign{\smallskip}
VIPERS & Input & 2000 & \\
 & Masking & 2000 & Mask value: 0 \\
 & Dense & 896 & Act.: ReLU \\
 & Dense & 1024 & Act.: ReLU \\
 & Dense & 512 & Act.: ReLU \\
 & Dense & 64 & Act.: ELU \\
 & Dense & 640 & Act.: ReLU \\
 & Dense & 512 & Act.: ReLU \\
 & Dense & 768 & Act.: ReLU \\
 & Dense & 2000 & \\
\hline
\end{tabular*}
\tablefoot{The activation functions (Act. value in the table) used are the exponential linear unit (ELU) and the rectified linear unit (ReLU) function.}
\end{table}
\FloatBarrier

\begin{table}[!h]
\caption{Architecture of the optimized CNNs applied to WEAVE (red and full spectra) and VIPERS data.}
\label{tab:cnn_architecture}
\centering
\begin{tabular*}{\columnwidth}{l @{\extracolsep{\fill}} l l l}
\hline\hline
\noalign{\smallskip}
Model & Type & Size & Additional Info \\\hline
\noalign{\smallskip}
WEAVE  & Input & 4020 & \\
red part & Masking & 4020 & Mask value: 0 \\
  & Conv & (4020, 32) & Kernel Size: 5 \\
 & &  & Act.: ReLU \\
 & MaxPooling & (2010, 32) & Pool Size: 2 \\
  & Conv & (2010, 160) & Kernel Size: 5 \\
 & &  & Act.: ReLU \\
 & MaxPooling & (1050, 160) & Pool Size: 2 \\
  & Conv & (1005, 64) & Kernel Size: 3 \\
 & &  & Act.: ReLU \\
 & MaxPooling & (335, 64) & Pool Size: 2 \\
  & Conv & (335, 256) & Kernel Size: 5 \\
 & &  & Act.: ELU \\
 & Flatten & 28416 &  \\
 & Dense & 384 & Act.: ReLU \\
 & Dense & 1024 & Act.: ReLU \\
 & Dense & 5000 & Act.: ReLU \\
  \hline
\noalign{\smallskip}
WEAVE  & Input & 5000 & \\
full spectra & Masking & 5000 & Mask value: 0 \\
  & Conv & (5000, 256) & Kernel Size: 5 \\
 & &  & Act.: ELU \\
 & MaxPooling & (2500, 256) & Pool Size: 2 \\
  & Conv & (2500, 192) & Kernel Size: 3 \\
 & &  & Act.: ReLU \\
 & MaxPooling & (833, 192) & Pool Size: 2 \\
  & Conv & (833, 64) & Kernel Size: 3 \\
 & &  & Act.: ReLU \\
 & MaxPooling & (277, 64) & Pool Size: 2 \\
  & Conv & (277, 160) & Kernel Size: 3 \\
 & &  & Act.: ELU \\
 & MaxPooling & (92, 162) & Pool Size: 2 \\
 & Dropout & (92, 162) & Drop rate: 0.2 \\
 & Flatten & 14720 &  \\
 & Dense & 512 & Act.: ReLU \\
 & Dense & 1024 & Act.: ReLU \\
 & Dense & 5000 & Act.: ReLU \\
  \hline
\noalign{\smallskip}
VIPERS  & Input & 2000 & \\
 & Masking & 2000 & Mask value: 0 \\
  & Conv & (2000, 32) & Kernel Size: 5 \\
 & &  & Act.: ReLU \\
 & MaxPooling & (1000, 32) & Pool Size: 2 \\
  & Conv & (1000, 160) & Kernel Size: 5 \\
 & &  & Act.: ReLU \\
 & MaxPooling & (500, 160) & Pool Size: 2 \\
  & Conv & (500, 64) & Kernel Size: 3 \\
 & &  & Act.: ReLU \\
 & MaxPooling & (166, 64) & Pool Size: 2 \\
  & Conv & (166, 256) & Kernel Size: 5 \\
 & &  & Act.: ELU \\
 & MaxPooling & (55, 256) & Pool Size: 2 \\
 & Flatten & 14080 &  \\
 & Dense & 384 & Act.: ReLU \\
 & Dense & 1024 & Act.: ReLU \\
 & Dense & 2000 & Act.: ReLU \\
 \hline
\noalign{\smallskip}
\end{tabular*}
    \tablefoot{The activation functions (Act. value in the table) used are the rectified linear unit (ReLU), exponential linear unit (ELU), and linear activation.}
\end{table}
\FloatBarrier

\begin{table*}[h]
\caption{Architecture of the optimized U-Nets applied to WEAVE and VIPERS data.}
\label{tab:unet}
\centering
\begin{tabular*}{\textwidth}{l @{\extracolsep{\fill}} l l l  l l l l} 
\hline\hline                        
Model & Type & Size & Additional Info & Model & Type & Size & Additional Info \\
\hline
WEAVE & Input & (5000, 1) &  & VIPERS & Input & (2000, 1) & \\
 & Masking & (5000, 1) & Mask value: 0 & & Masking & (2000, 1) & Mask value: 0\\
 & Conv & (5000, 16) & Kernel Size: 5 & & Conv & (2000, 32) & Kernel Size: 5 \\
 & &  & Act.: ReLU & & & & Act.: ReLU \\
 & Conv & (5000, 16) & Kernel Size: 5 & & Conv & (2000, 32) & Kernel Size: 5 \\
 & &  & Act.: ReLU & & & & Act.: ReLU \\
 & MaxPooling & (2500, 16) & Pool Size: 2 & & MaxPooling & (1000, 32) & Pool Size: 2 \\
 & Conv & (2500, 32) & Kernel Size: 5 & & Conv & (1000, 128) & Kernel Size: 5 \\
 & &  & Act.: ReLU & & & & Act.: ReLU \\
 & MaxPooling & (1250, 32) & Pool Size: 2 & & Conv & (1000, 128) & Kernel Size: 5 \\
 & Conv & (1250, 128) & Kernel Size: 5 & & & & Act.: ReLU \\
 & &  & Act.: ReLU & & MaxPooling & (500, 128) & Pool Size: 2 \\
 & Conv & (1250, 128) & Kernel Size: 5 & & Conv & (500, 128) & Kernel Size: 5 \\
 & &  & Act.: ReLU & & & & Act.: ReLU \\
 & MaxPooling & (625, 128) & Pool Size: 2 & & Conv & (500, 128) & Kernel Size: 5 \\
 & Conv & (625, 128) & Kernel Size: 5 & & & & Act.: ReLU \\
 & &  & Act.: ReLU & & MaxPooling & (250, 128) & Pool Size: 2 \\
 & Conv & (625, 128) & Kernel Size: 5 & & Conv & (250, 512) & Kernel Size: 5 \\
 & &  & Act.: ReLU & & & & Act.: ReLU \\
 & Conv & (1250, 128) & Kernel Size: 2 & & Conv & (250, 512) & Kernel Size: 5 \\
 & Transpose & & Act.: ReLU & & & & Act.: ReLU \\
 & Concatenate & (1250, 256) & & & Conv & (500, 128) & Kernel Size: 5 \\
 & Conv & (1250, 128) & Kernel Size: 5 & & Transpose & & Act.: ReLU \\
 & &  & Act.: ReLU & & Concatenate & (500, 256) & \\
 & Conv & (1250, 128) & Kernel Size: 5 & & Conv & (500, 128) & Kernel Size: 5 \\
 & &  & Act.: ReLU & & & & Act.: ReLU \\ 
 & Conv & (2500, 32) & Kernel Size: 2 & & Conv & (500, 128) & Kernel Size: 5 \\
 & Transpose & & Act.: ReLU & & & & Act.: ReLU \\
 & Concatenate & (2500, 64) & & & Conv & (1000, 128) & Kernel Size: 2 \\
 & Conv & (2500, 32) & Kernel Size: 5 & & Transpose & & Act.: ReLU \\
 & &  & Act.: ReLU & & Concatenate & (1000, 256) & \\
 & Conv & (2500, 32) & Kernel Size: 5 & & Conv & (1000, 128) & Kernel Size: 5 \\
 & &  & Act.: ReLU & & & & Act.: ReLU \\
 & Conv & (5000, 16) & Kernel Size: 2 & & Conv & (2000, 32) & Kernel Size: 2 \\
 & Transpose & & Act.: ReLU & & Transpose & & Act.: ReLU \\
 & Concatenate & (5000, 32) & & & Concatenate & (2000, 64) & \\
 & Conv & (5000, 16) & Kernel Size: 5 & & Conv & (2000, 128) & Kernel Size: 5 \\
 & &  & Act.: ReLU & & & & Act.: ReLU \\
 & Conv & (5000, 16) & Kernel Size: 5 & & Conv & (2000, 128) & Kernel Size: 5 \\
 & &  & Act.: ReLU & & & & Act.: ReLU \\
 & Conv & (5000, 1) & Kernel Size: 1 & & Conv & (2000, 1) & Kernel Size: 1 \\
\hline
\end{tabular*}
\tablefoot{The activation function (Act. value in the table) used is the rectified linear unit (ReLU) function.}
\end{table*}
\FloatBarrier




\section{Fit example: autoencoders and CNNs}\label{app:fit}

\begin{figure*}[!h]
    \centering
    \resizebox{\hsize}{!}{\includegraphics{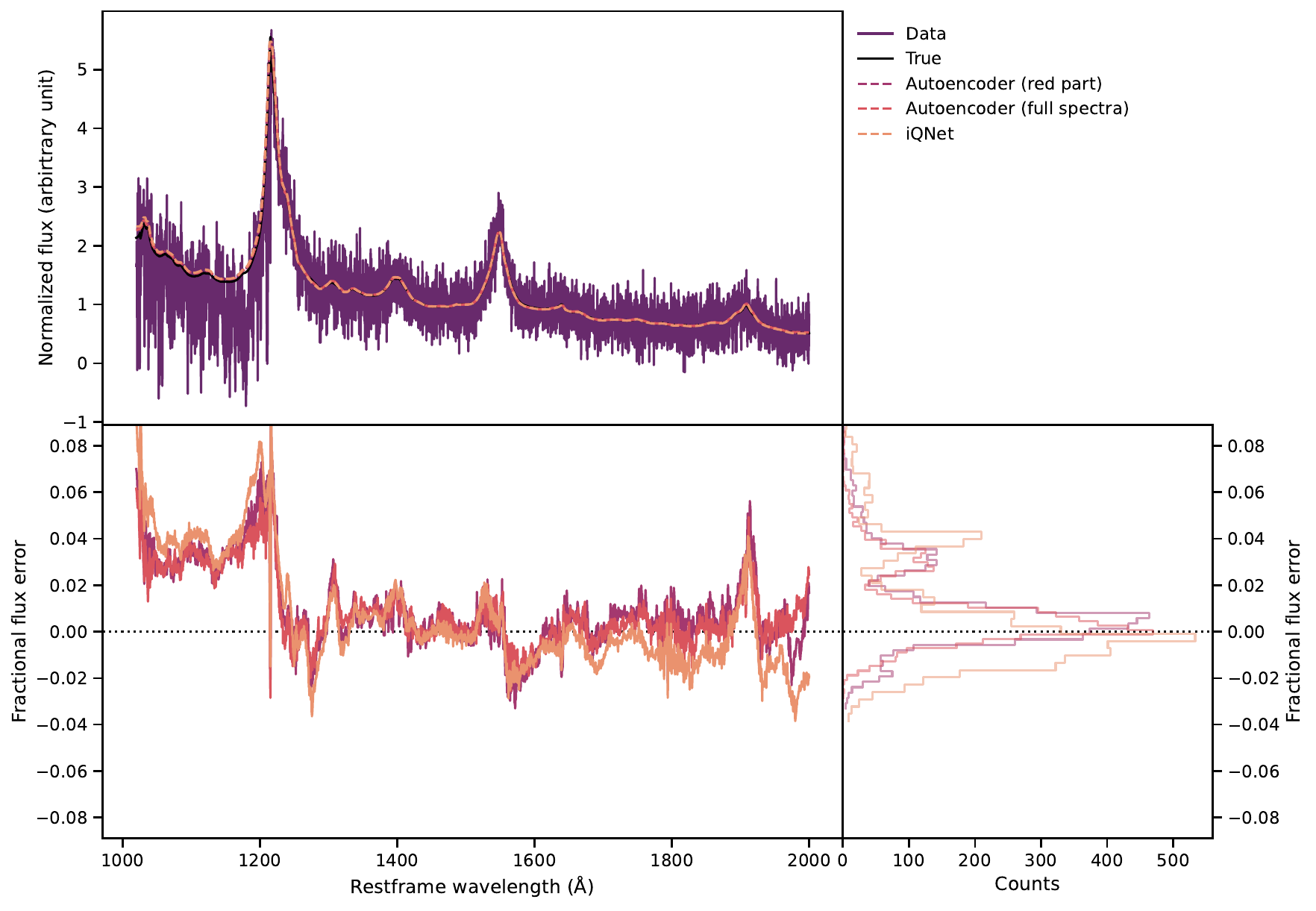}\includegraphics{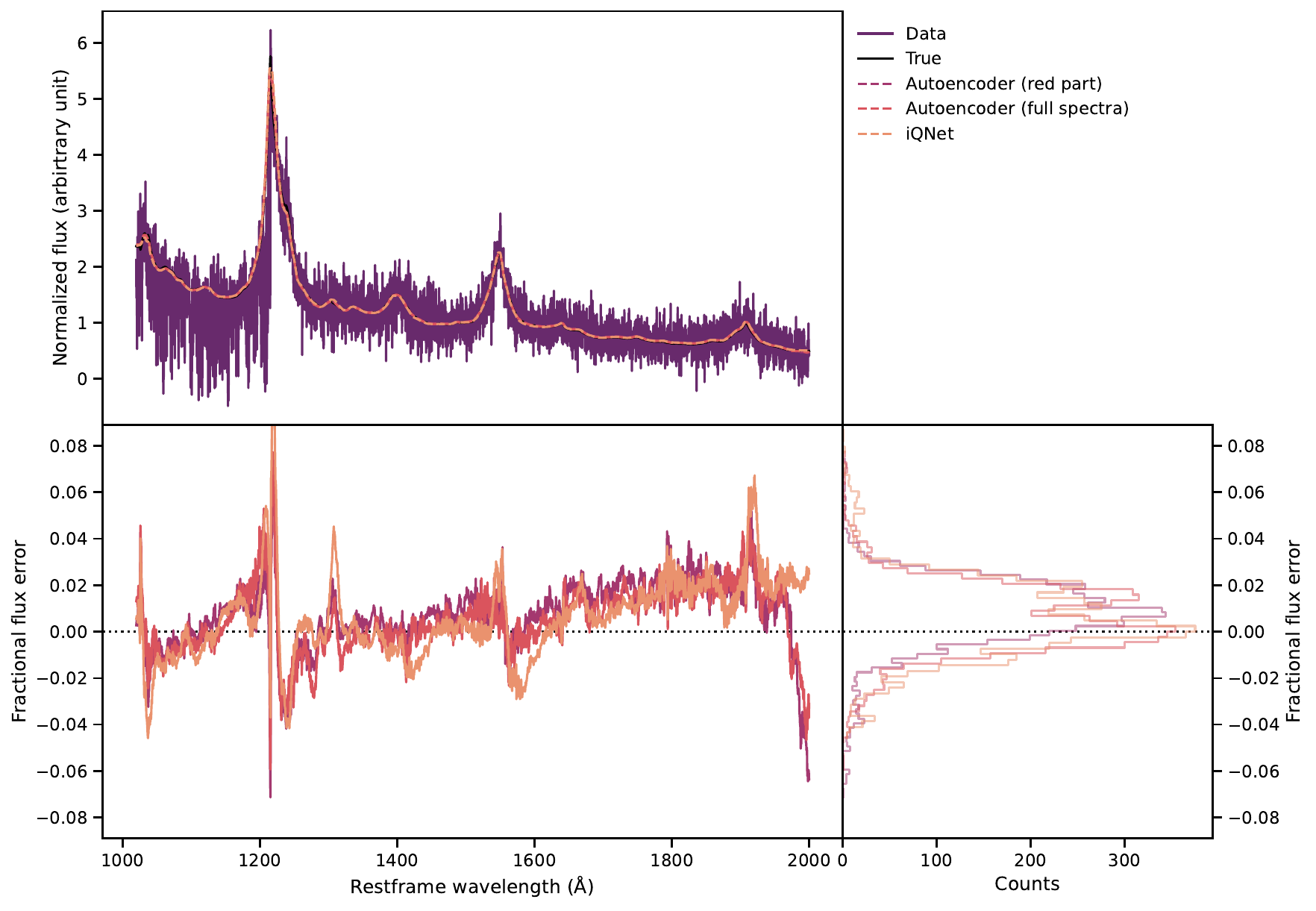}}
    \resizebox{\hsize}{!}{\includegraphics{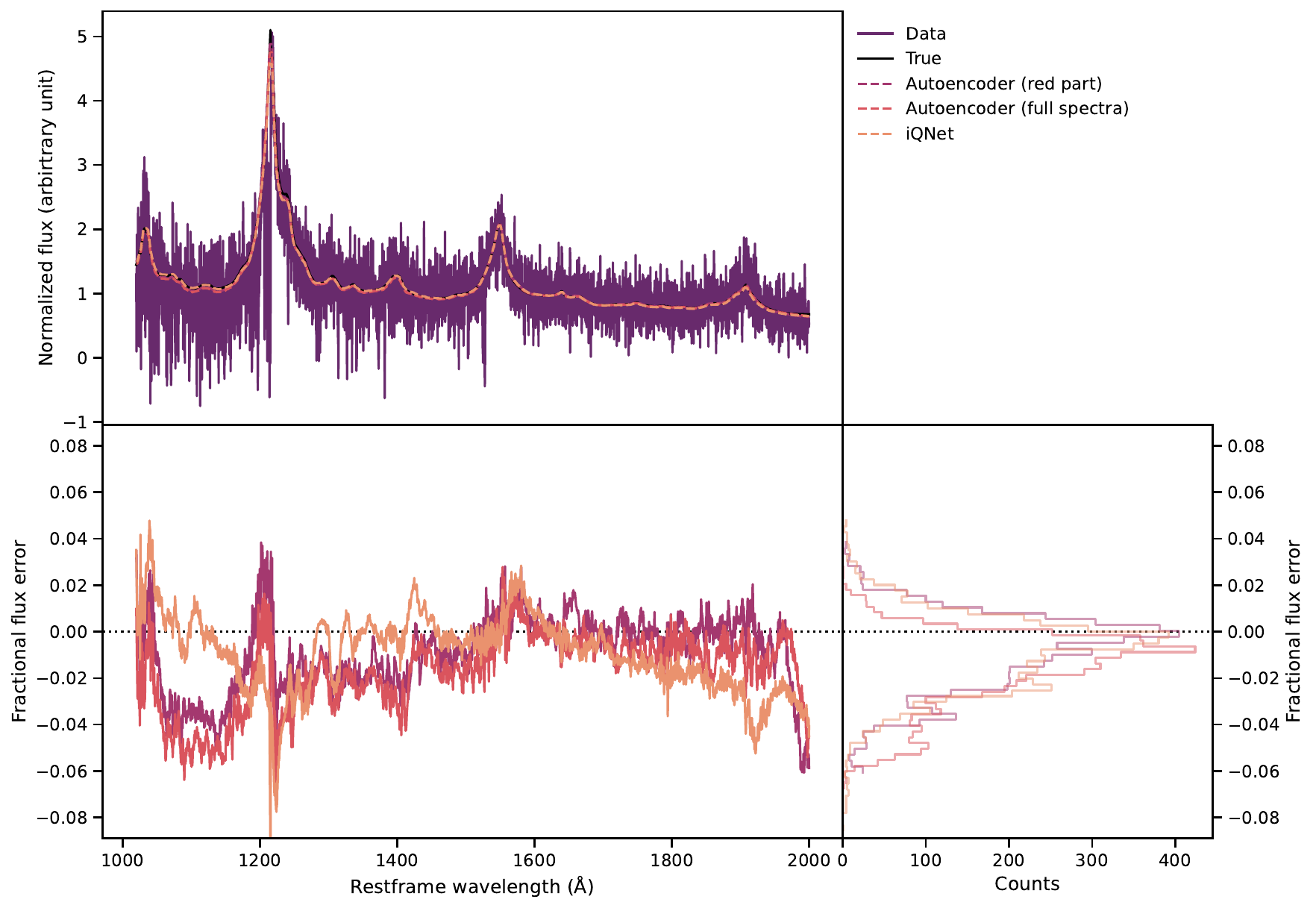}\includegraphics{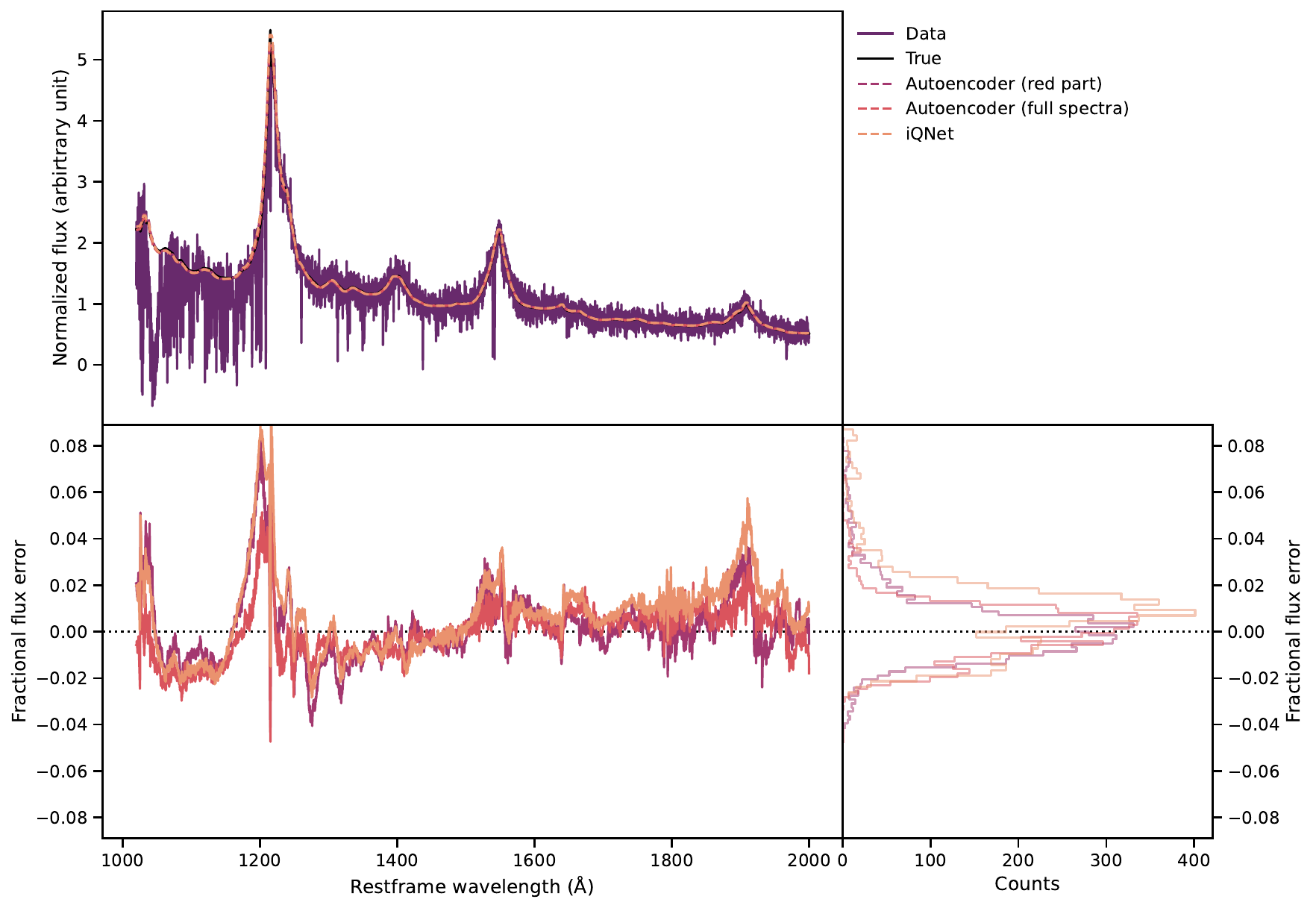}}
    \resizebox{\hsize}{!}{\includegraphics{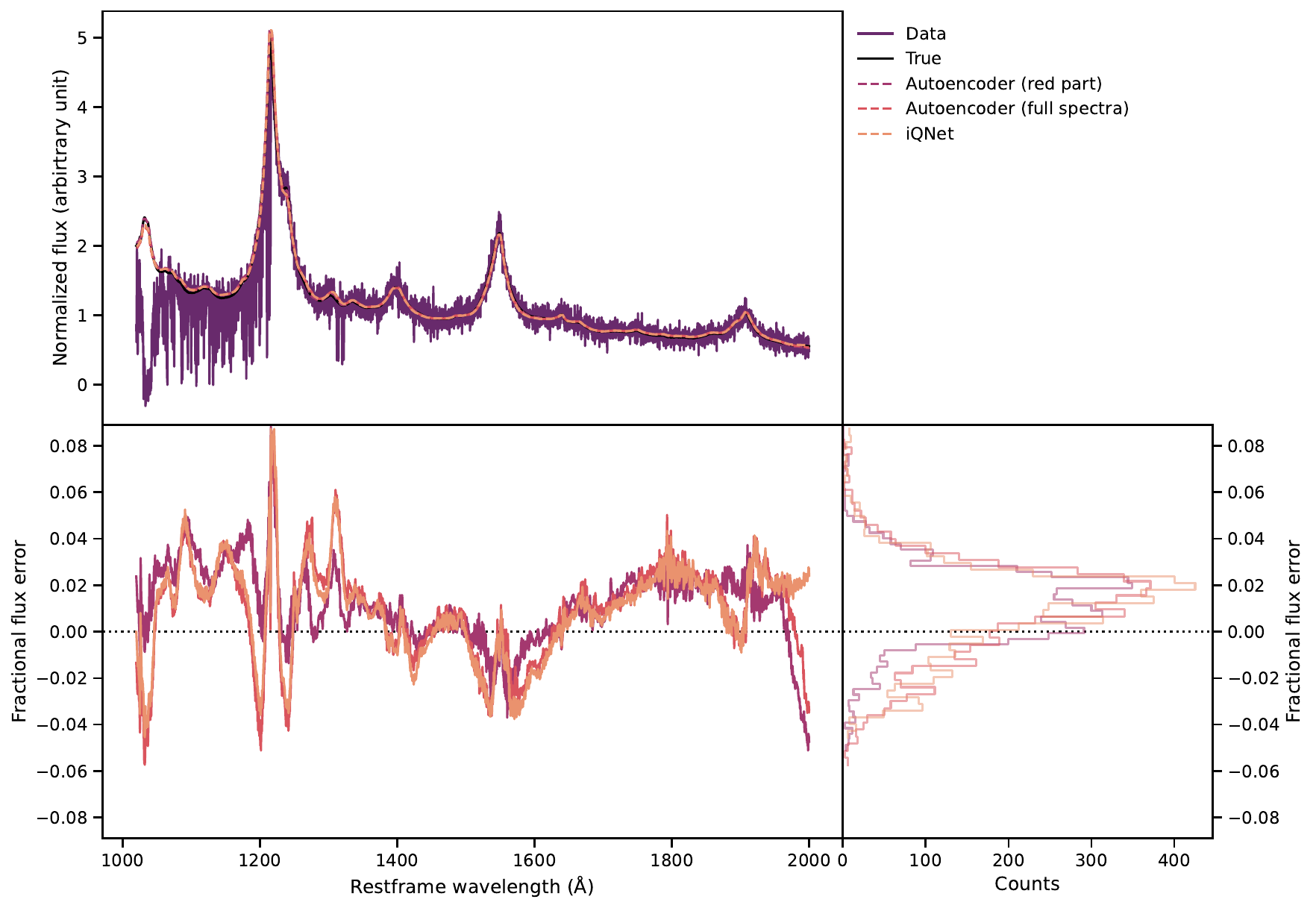}\includegraphics{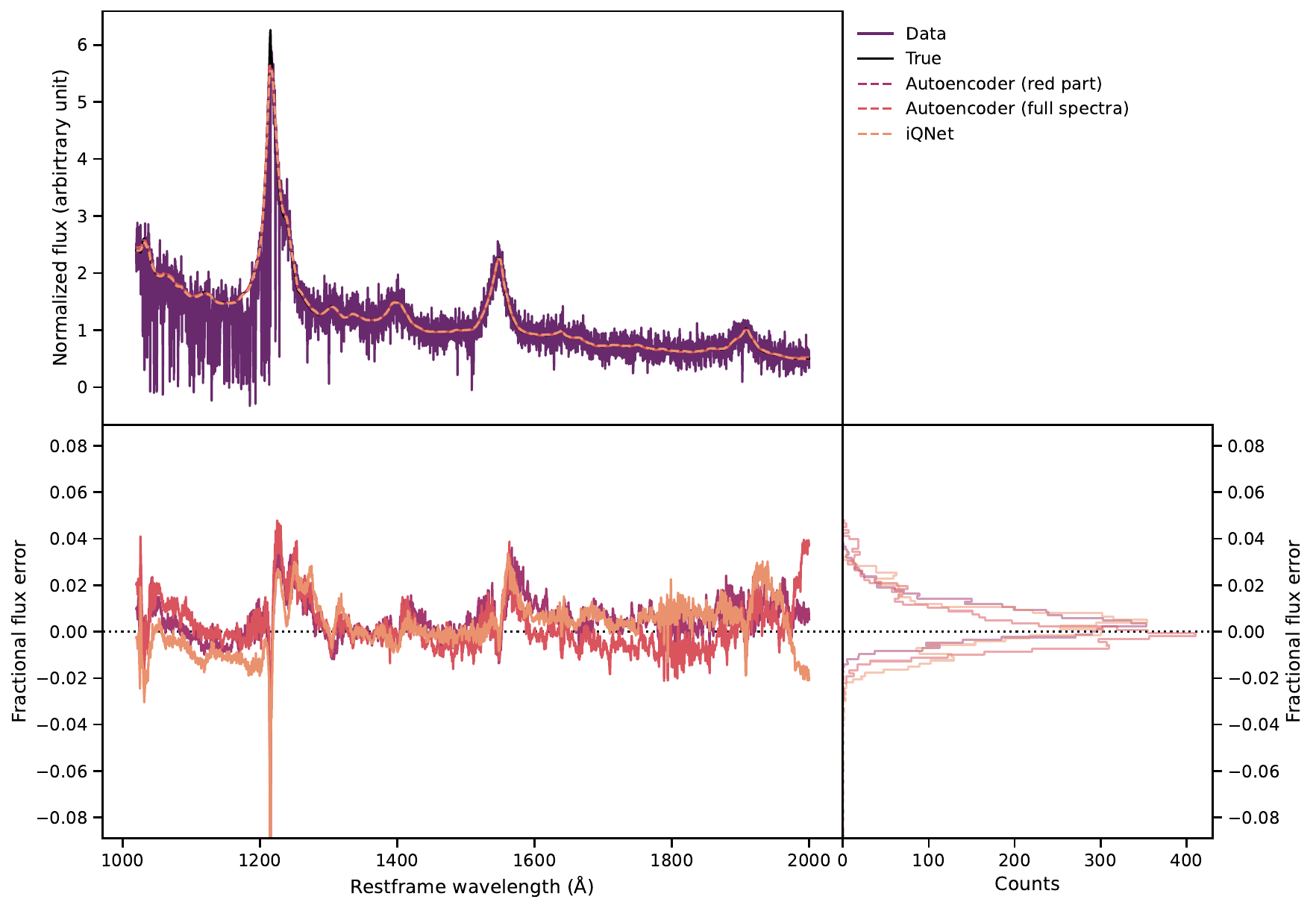}}
    \caption{Example spectra of quasars. The top panel of each plot shows the data, the true continuum, and the fits of different runs of the autoencoders. The bottom panel shows the FFE as a function of the rest-frame wavelength for different runs of the autoencoders and their distributions.}
    \label{fig:aut_fit_weave}
\end{figure*}

\begin{figure*}[!h]
    \centering
    \resizebox{\hsize}{!}{\includegraphics{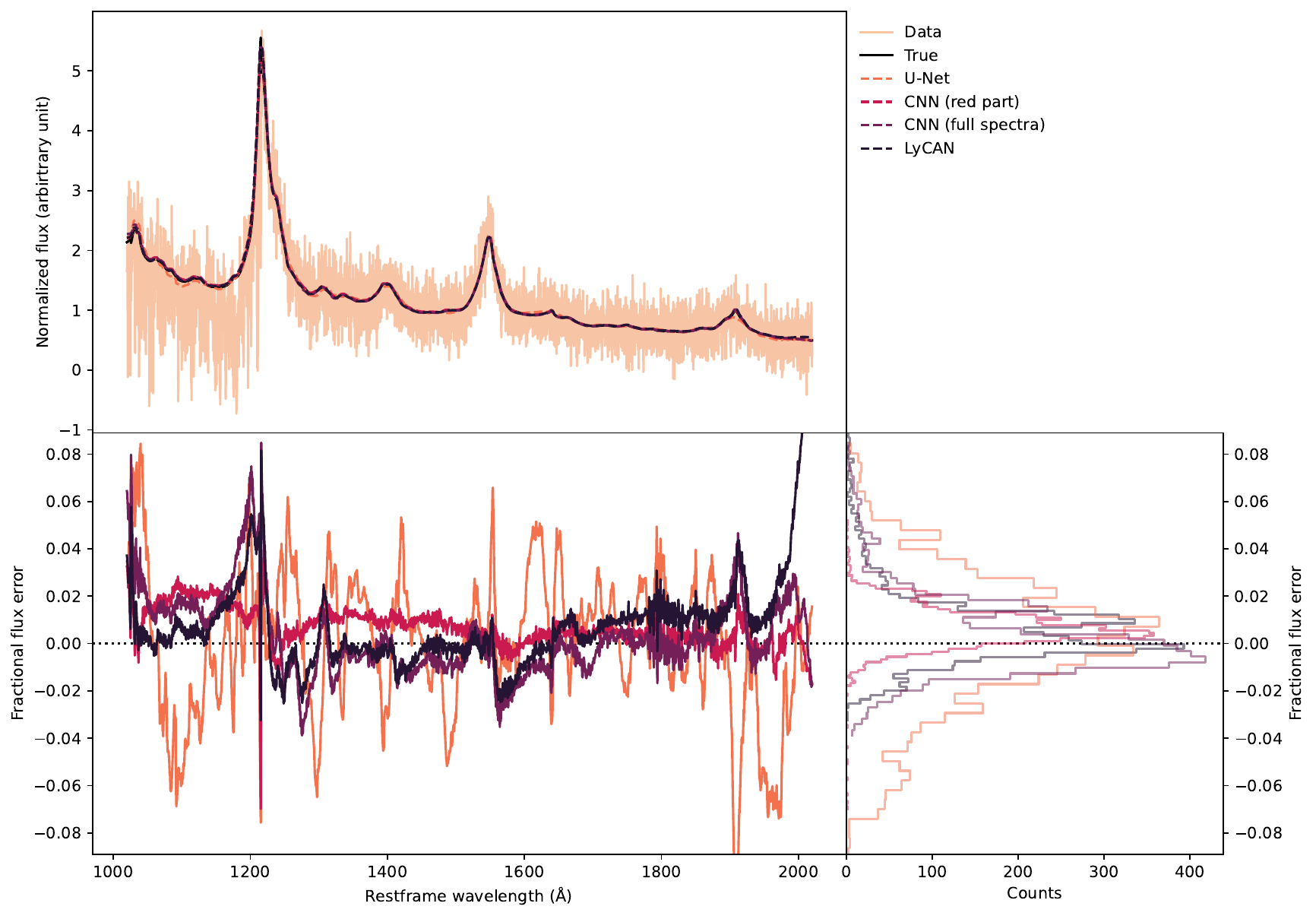}\includegraphics{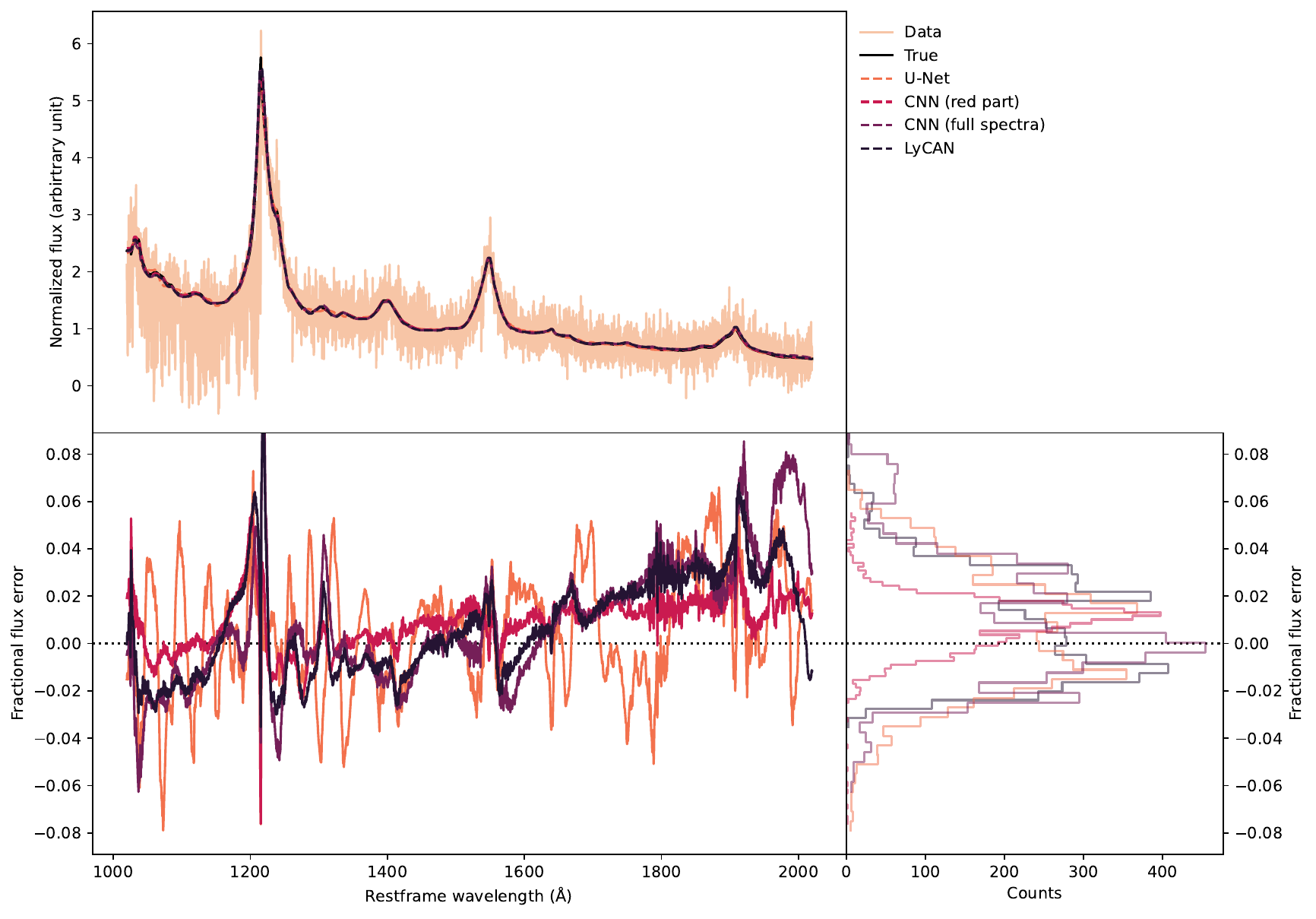}}
    \resizebox{\hsize}{!}{\includegraphics{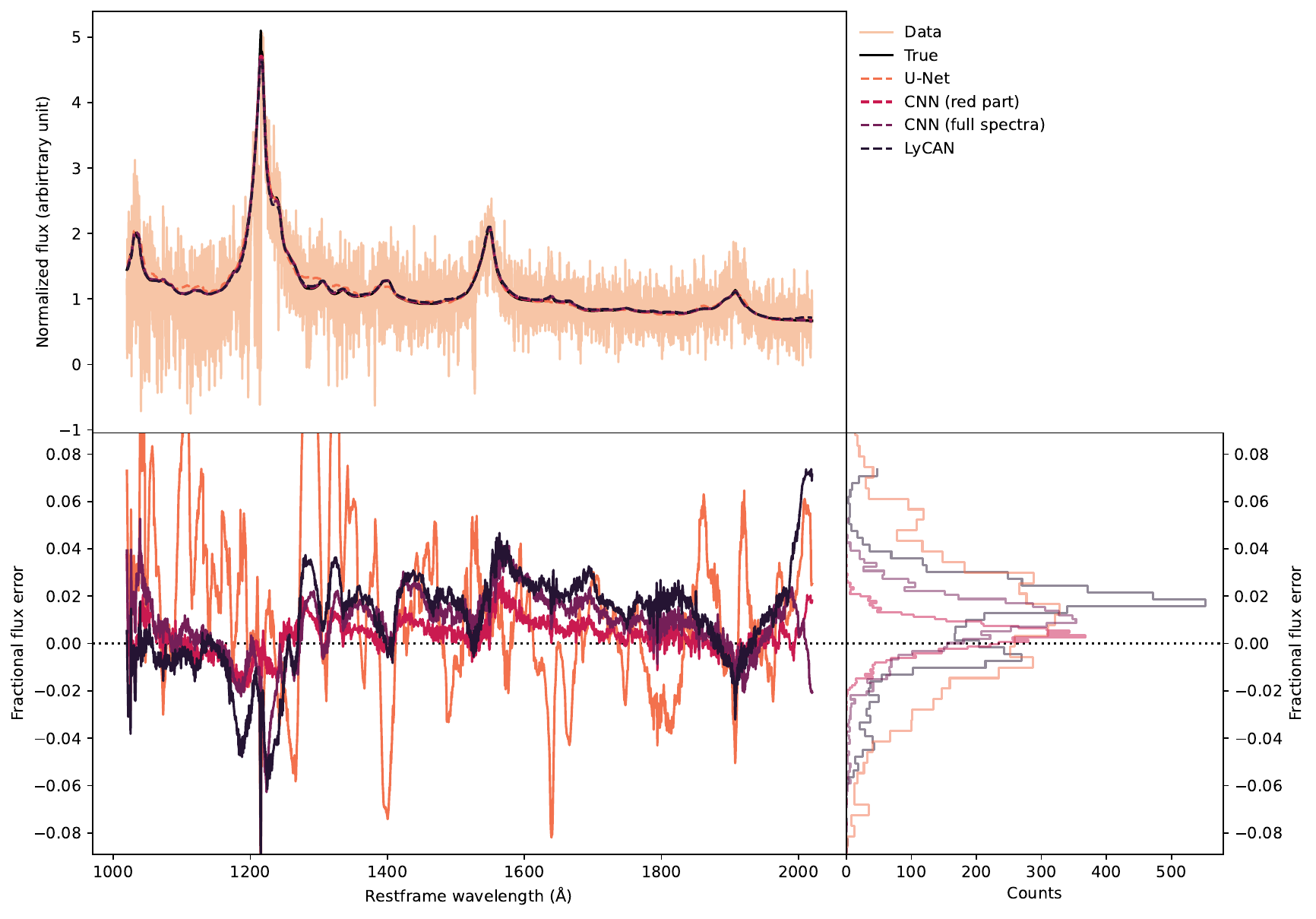}\includegraphics{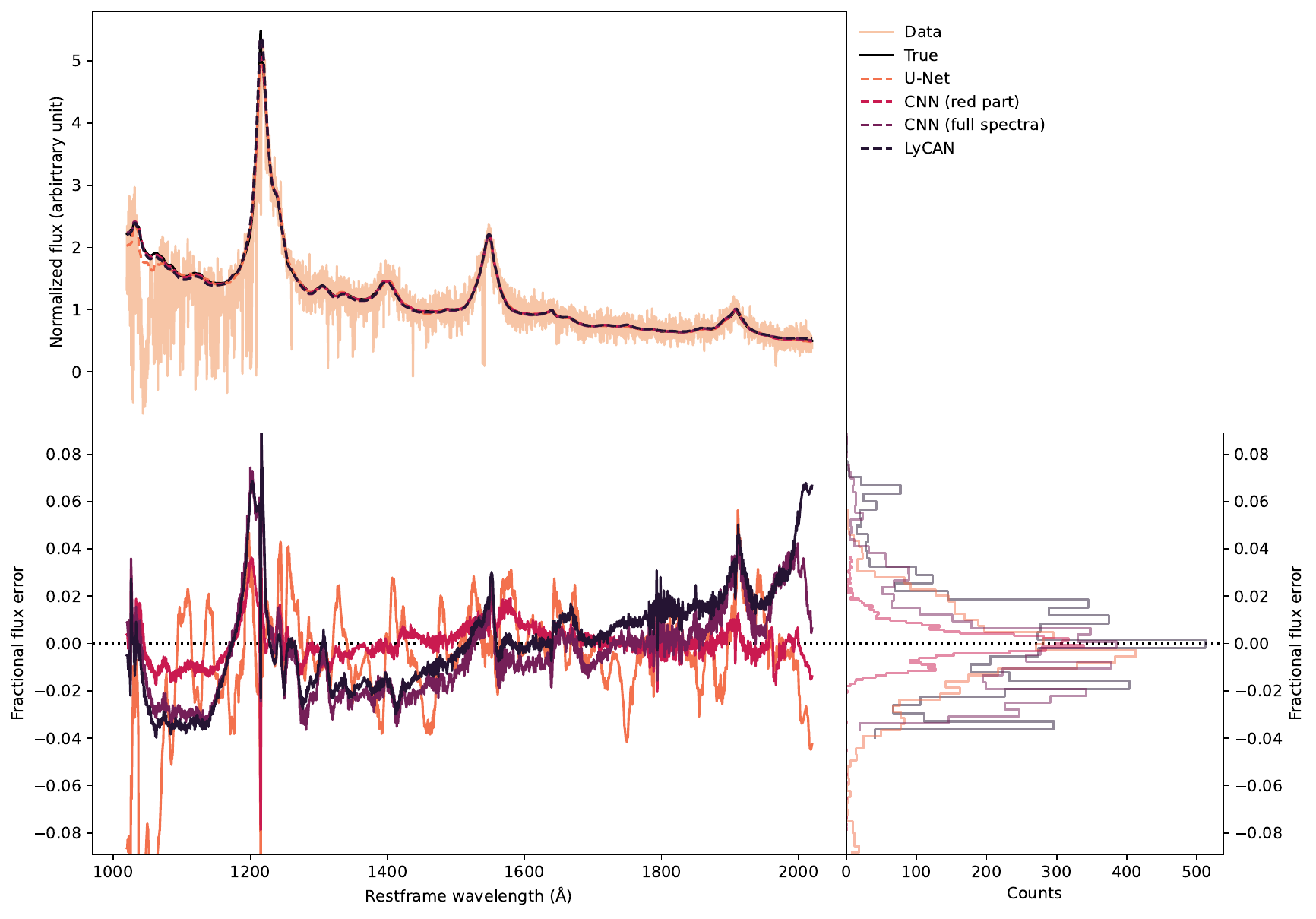}}
    \resizebox{\hsize}{!}{\includegraphics{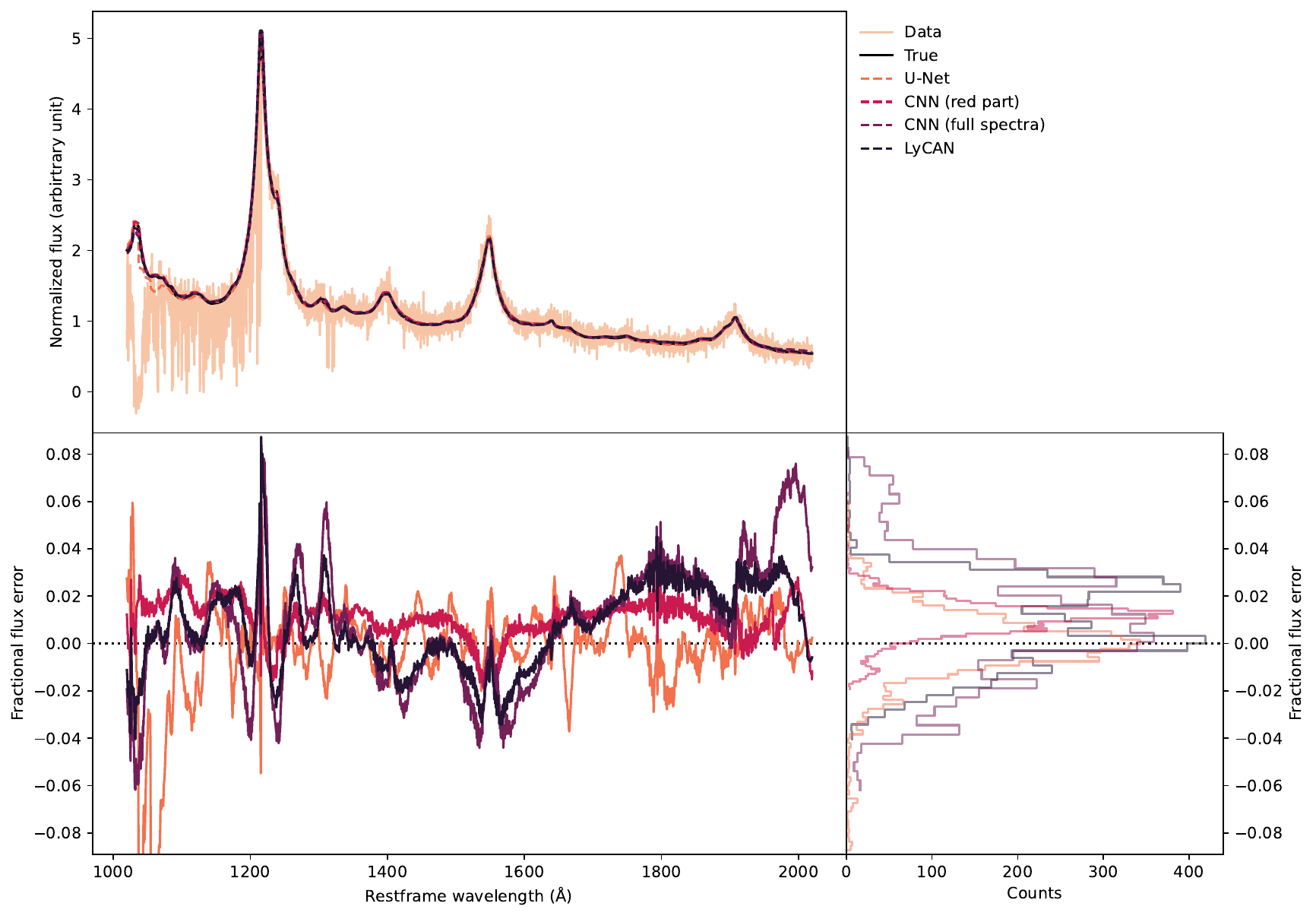}\includegraphics{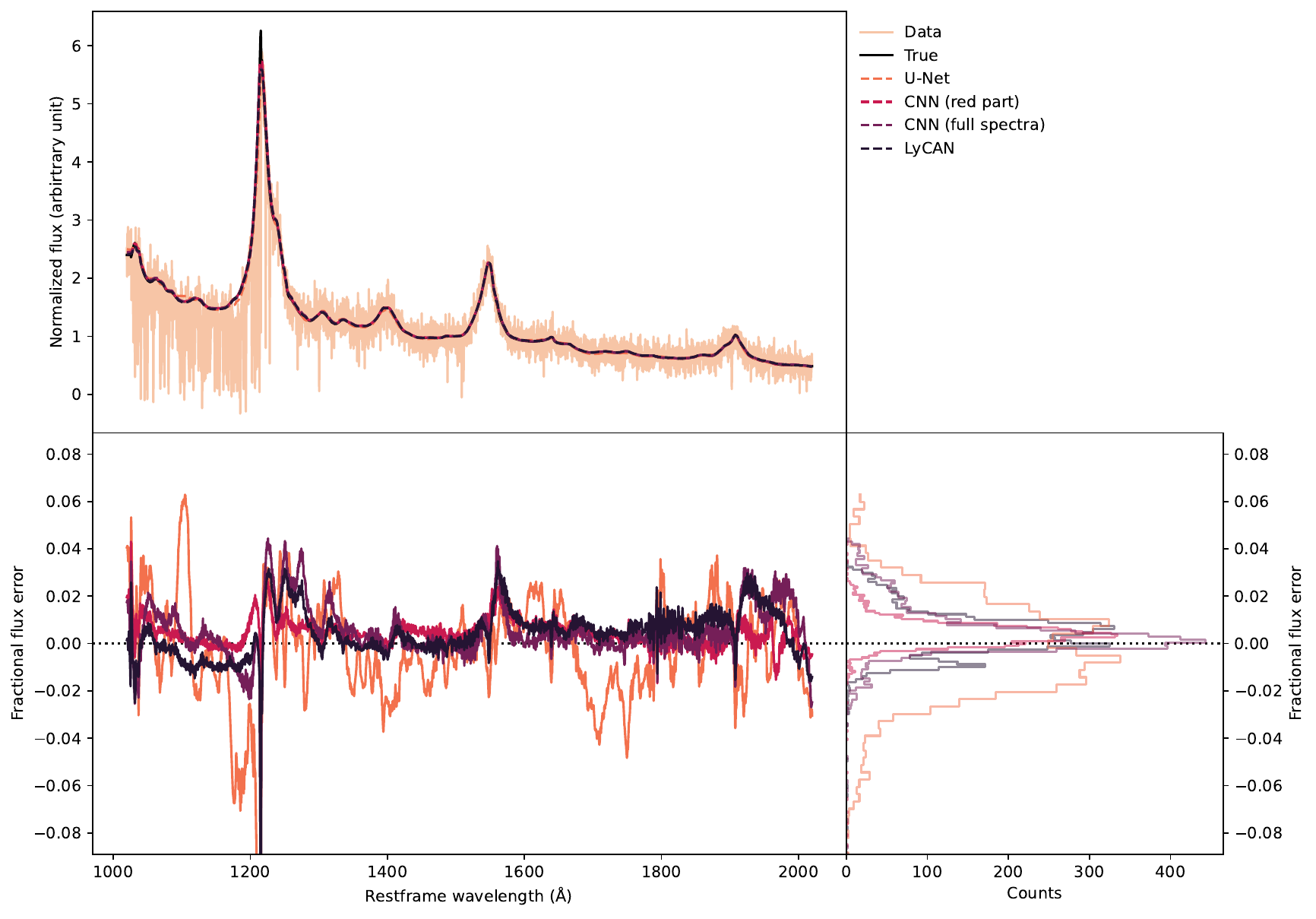}}
    \caption{Example spectra of quasars. The top panel of each plot shows the data, the true continuum, and the fits of different runs of the CNNs. The bottom panel shows the FFE as a function of the rest-frame wavelength for different runs of the autoencoders and their distributions. The shown spectra are the same as Fig.~\ref{fig:aut_fit_weave}.}
    \label{fig:cnn_fit_weave}
\end{figure*}

Fig.~\ref{fig:aut_fit_weave} shows examples of spectra for the case of giving to the optimized autoencoder only the red part of the spectra or the whole spectra as input and iQNet.
Fig.~\ref{fig:cnn_fit_weave} shows examples of spectra using U-Net and Lycan.
\FloatBarrier
\FloatBarrier


\end{appendix}

\end{document}